\documentclass[lettersize,journal]{IEEEtran}
\usepackage{amsmath,amsfonts}
\usepackage{algorithmic}
\usepackage{array}
\usepackage{textcomp}
\usepackage{stfloats}
\usepackage{url}
\usepackage{verbatim}
\usepackage{graphicx}

\usepackage{txfonts}
\usepackage{cite}
\usepackage{comment}
\usepackage{bm}
\usepackage{hyperref}
\usepackage{cleveref}
\usepackage{xspace}
\usepackage{array}
\usepackage{ulem}
\usepackage{makecell}
\usepackage{adjustbox}
\usepackage{tabularx}
\usepackage{rotating}
\usepackage{booktabs}
\usepackage{threeparttable}
\usepackage{enumitem}
\usepackage{bbding}
\usepackage{multirow}
\usepackage{xcolor}
\usepackage{colortbl}
\usepackage[ margin=6pt]{subfig}

\newcolumntype{C}{>{\centering\arraybackslash}X}

\newcommand{\sihan}[1][\textcolor{black}]{#1}

\newcommand{\mname}{LAN\xspace}

\newcommand{\moduleone}{Activity Sequence Modeling\xspace}
\newcommand{\moduletwo}{Activity Graph Learning\xspace}
\newcommand{\modulethree}{Anomaly Score Prediction\xspace}

\def\BibTeX{{\rm B\kern-.05em{\sc i\kern-.025em b}\kern-.08em
    T\kern-.1667em\lower.7ex\hbox{E}\kern-.125emX}}
\usepackage{balance}
\begin{document}
\title{LAN: Learning Adaptive Neighbors for Real-Time Insider Threat Detection}
\author{Xiangrui~Cai, Yang~Wang, Sihan~Xu\textsuperscript{*}, Hao~Li, Ying~Zhang, Zheli~Liu, Xiaojie~Yuan
\thanks{This work is supported by the National Key R\&D Program of China (2022YFB3103202) and the National Science Foundation of China (62372252, U22B2048, and 62272250).}
\thanks{Yang Wang, Sihan Xu, Zheli Liu, Xiaojie Yuan are with the Key Laboratory of Data and Intelligent System Security, Ministry of Education, China and the College of Cyber Science, Nankai University, Tianjin 300350, China (e-mail: wangyang@dbis.nankai.edu.cn, xusihan@nankai.edu.cn, liuzheli@nankai.edu.cn, yuanxj@nankai.edu.cn). 
Xiangrui Cai, Ying Zhang is with the College of Computer Science, Nankai University, Tianjin 300350, China (e-mail: caixr@nankai.edu.cn, yingzhang@nankai.edu.cn). Hao Li is with the Science and Technology on Communication Networks Laboratory, Shijiazhuang 050081, China (e-mail: cuclihao@cuc.edu.cn).}
\thanks{\textsuperscript{*}Corresponding author.}
}

\markboth{}
{Xiangrui Cai \MakeLowercase{\textit{et al.}}: \mname: Learning Adaptive Neighbors for Real-Time
Insider Threat Detection}

\maketitle

\begin{abstract}
Enterprises and organizations are faced with potential threats from insider employees that may lead to serious consequences. Previous studies on insider threat detection (ITD) mainly focus on detecting abnormal users or abnormal time periods (e.g., a week or a day). However, a user may have hundreds of thousands of activities in the log, and even within a day there may exist thousands of activities for a user, requiring a high investigation budget to verify abnormal users or activities given the detection results. On the other hand, existing works are mainly post-hoc methods rather than real-time detection, which can not report insider threats in time before they cause loss. In this paper, we conduct the first study towards real-time ITD at activity level, and present a fine-grained and efficient framework \mname. Specifically, \mname simultaneously learns the temporal dependencies within an activity sequence and the relationships between activities across sequences with graph structure learning. Moreover, to mitigate the data imbalance problem in ITD, we propose a novel hybrid prediction loss, which integrates self-supervision signals {from normal activities} and supervision signals from abnormal activities into a unified loss for anomaly detection. We evaluate the performance of \mname on two widely used datasets, i.e., CERT r4.2 and CERT r5.2. 
Extensive and comparative experiments demonstrate the superiority of \mname, outperforming 9 state-of-the-art baselines by at least 9.92\% and 6.35\% in AUC for real-time ITD on CERT r4.2 and r5.2, respectively. Moreover, \mname can be also applied to post-hoc ITD, surpassing 8 competitive baselines by at least 7.70\% and 4.03\% in AUC on two datasets. Finally, the ablation study, parameter analysis, and compatibility analysis evaluate the impact of each module and hyper-parameter in \mname. The source code can be obtained from \url{https://github.com/Li1Neo/LAN}.
\end{abstract}

\begin{IEEEkeywords}
Insider threat detection, activity-level detection, real-time detection, graph structure learning, class imbalance.
\end{IEEEkeywords}

\begin{figure}[t]
     \centering
     	\centering
	\subfloat[ITD illustration]{
		\begin{minipage}[b]{0.92\linewidth}
			\includegraphics[width=1\textwidth]{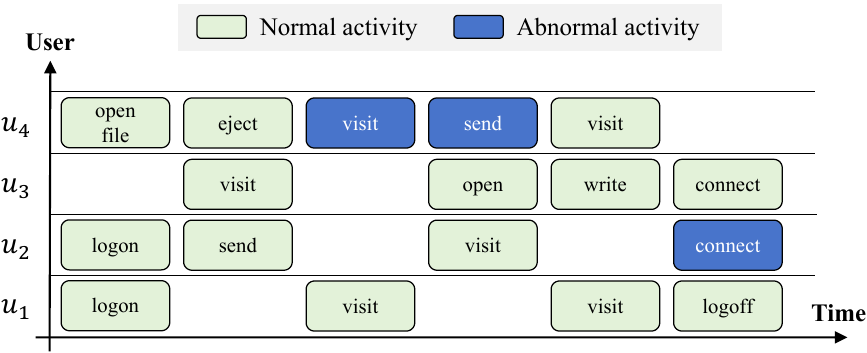} 
		\end{minipage}
		\label{fig:fig1a_illustration}
	}
	\\ 
	\subfloat[Post-hoc ITD]{
		\begin{minipage}[b]{0.45\linewidth}
			\includegraphics[width=1\textwidth]{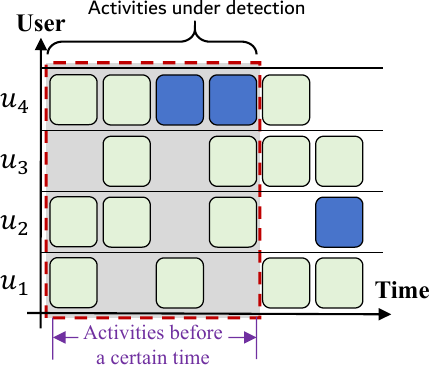} 
		\end{minipage}
		\label{fig:fig1b_PHITD}
	}
    \subfloat[Real-time ITD]{
        \begin{minipage}[b]{0.45\linewidth}
        \includegraphics[width=1\textwidth]{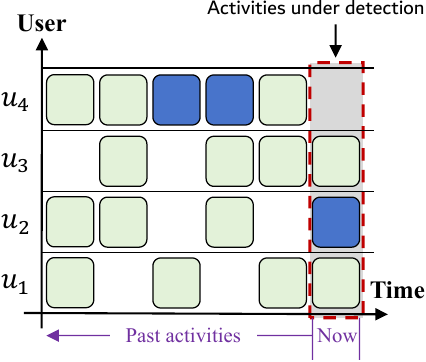}
        \end{minipage}
    \label{fig:fig1c_RTITD}
    }
	\caption{(a) Illustration of Activity-level ITD. 
 Activity-level ITD aims at discovering \textbf{abnormal activities} inside the system (shown in {blue boxes}). (b) Post-hoc ITD. It is usually deployed for retrospective detection, discovering abnormal activities in a past period. (c) Real-time ITD. It is usually applied to detect abnormality of a current activity.}
	\label{fig:fig1}
\end{figure}

\section{Introduction}
\IEEEPARstart{M}{odern} information systems are vulnerable to attacks from insider employees who have authorized access to {system, data, or {internal} network}~\cite{silowash2012common}.
Such insider threats may corrupt the integrity, confidentiality, and availability of systems~\cite{costa2016insider}. 
From 2020 to 2022, the number of insider threat incidents had increased by 44\%, resulting in the average costs rising by 34\%, from \$11.45 million to \$15.38 million~\cite {proofpoint2022}.
Due to the destructive effects of insider threats, much effort has been devoted to Insider Threat Detection (ITD) to prevent from unpredictable impact brought by insider threats.
Specifically, existing works on ITD can be grouped into three categories, i.e., the feature engineering-based methods{~\cite{le2020analyzing, yuan2021time}}, the sequence-based methods{~\cite{tuor2017deep,yuan2019insider,vinay2022contrastive}}, and the graph-based methods{~\cite{jiang2019anomaly,DBLP:journals/tifs/LiLJLYGY23, DBLP:conf/ccs/LiuWZJXM19}}. 
The first group extracts features such as the number of file accesses after work hours~\cite{le2020analyzing}, and then trains a machine learning model to detect insider threats.
{The second group collects user activity sequence data for each time period (e.g., a day~\cite{tuor2017deep} or a session~\cite{vinay2022contrastive}), and then trains a sequence-based model to predict whether a specific time period is abnormal or not.}

Inspired by graph-based methods in the field of intrusion detection~\cite{wang2022threatrace,wang2022wrongdoing}, the third group proposes constructing a graph that incorporates relationships between users ~\cite{jiang2019anomaly, DBLP:journals/tifs/LiLJLYGY23} or activities ~\cite{DBLP:conf/ccs/LiuWZJXM19} across different sequences, to detect insider threats.

Despite the progress, existing approaches are faced with two problems, which limit their practical application in real-world ITD. \textbf{First}, most studies on ITD focus on detecting either abnormal users~\cite{DBLP:journals/tifs/LiLJLYGY23} or abnormal time periods {(e.g., a week or a day}). {However, a user may have hundreds of thousands of activities in the log, and even within a day there may exist thousands of activities for a user. As a result, these approaches are too course-grained to be adopted, requiring a high investigation budget to verify the abnormal users or activities from the detection results.}
Actually, as shown in \Cref{fig:fig1a_illustration}, an insider threat accident can be directly reflected by a set of abnormal activities of a user, which are more fine-grained and accurate.
\textbf{Second}, existing works are mainly post-hoc methods rather than real-time detection. {However, it is more desirable to report insider threats in time before they cause loss~\cite{hu2023towards}.}
~\Cref{fig:fig1b_PHITD} and ~\Cref{fig:fig1c_RTITD} illustrate \textit{post-hoc} ITD and \textit{real-time} ITD, respectively.
It can be seen that \textit{post-hoc} ITD identifies abnormal users or activities after an insider threat accident occurred and all the related data have been collected (e.g., logs during the past year). In contrast, \textit{real-time} ITD monitors the entire system and detects ITD at runtime. Once an insider threat accident occurs, it is identified and reported promptly, which can effectively prevent the organization or enterprise from financial loss.

{To address the aforementioned issues, this paper proposes a novel activity modeling framework named \mname for real-time ITD.
\mname not only excels in runtime insider threat detection but also offers an activity-level solution to detect anomalies in a fine-grained way.}
Specifically, \mname consists of three modules, i.e., \moduleone, \moduletwo, and \modulethree.
{Given a user activity sequence, \moduleone module utilizes a sequence encoder with attentive aggregation operation to model the temporal dependencies and obtain the representation of each activity.}
{To incorporate relationships across different activity sequences, we design the \moduletwo module to automatically learn activity graphs.}
{Then,} the \modulethree \  module employs a graph neural network to aggregate neighbors in the {activity graph, so as to enhance the representation of the current activity.} 
{Finally, the anomaly score is obtained by calculating the probability of the next behavior occurrence based on the enhanced activity representation.} 

Moreover, due to the fine-grained detection, the data imbalance between normal and abnormal samples for activity-level ITD is even worse than the problem for the user- or period-level ITD. This imbalance leads the model to lean towards predicting normal activities, thereby reducing the overall detection rate.
To alleviate this issue, we introduce a novel hybrid loss, which integrates both supervision of abnormal activities and self-supervision of the normal activity sequence {simultaneously}.

We evaluated \mname on two widely-used public datasets (i.e., CERT r4.2 and CERT r5.2) and compared \mname against 9 baselines for real-time ITD. 
We also applied \mname for post-hoc ITD with slight modifications and compared it with 8 state-of-the-art approaches.
The experimental results demonstrate that  \mname outperforms all baselines with average improvements of 9.53\% and 6.55\% in AUC for real-time ITD and post-hoc ITD, respectively. 
We further conduct ablation studies and parameter analysis to evaluate the effectiveness of each module, hyper-parameter, and the proposed hybrid prediction loss for the data imbalance problem in real-time ITD.

{In summary, we made the following novel contributions:
\begin{itemize}
    \item To our best knowledge, we conduct the first study towards {\textit{activity-level} \textit{real-time}} insider threat detection. 
    Specifically, we present a fine-grained and {efficient} framework named \mname, which employs graph structure learning to
    learn user activity graph adaptively, avoiding the bias introduced by manual graph construction.
    \item To alleviate the significant imbalance between normal and abnormal activities, we propose a novel hybrid prediction loss, which integrates self-supervision signals {from normal activities} and supervision signals from abnormal activities into a unified loss for anomaly detection.
    \item {
    Extensive and comparative experiments demonstrate the superiority of \mname, outperforming 9 state-of-the-art baselines by at least 9.92\% and 6.35\% in AUC for real-time ITD on CERT r4.2 and r5.2, respectively. Moreover, \mname can be also applied to post-hoc ITD, surpassing 8 competitive baselines by at least 7.70\% and 4.03\% in AUC on two datasets.}
\end{itemize}
}

\section{Preliminaries}\label{sec:pre}

{In this section, we formulate the problem of activity-level ITD, including post-hoc ITD and real-time ITD.}
{
Given an information system that can be accessed by $N$ users, we denote the user set by $\mathcal{U}=\{u_i \mid i\in\mathbb{N}^+, i\le N\}$.
The system records the activity sequence of each user chronologically, e.g., \texttt{logon}, \texttt{visit website}, and {\texttt{copy file}}.
We use $\mathcal{A}=\{a_i \mid i\in\mathbb{N}^+, i\le M\}$ to denote the set of all activities inside the system, where $M$ is the number of unique activities.
An activity sequence of a user $u\in \mathcal{U}$ is denoted by $S^u=\{a^{u}_1, a^u_2, \dots, a^u_{n_u}\}$, where $a^{u}_i\in \mathcal{A} (i=1, 2, \dots, n_u)$ and $n_u$ is the activity sequence length for user $u$.
The activities in $S^{u}$ are arranged in chronological order.
Note that the activity sequence length (i.e., $n_u$) could vary for different users. 
The activity sequences of all users form the whole sequence set $\mathcal{D}=\{S^u \mid u\in \mathcal{U}\}$ within the system.
}

{
\underline{Post-Hoc Insider Threat Detection}. 
Post-hoc ITD aims at retrospectively determining whether insider threats once occurred.
It collects user activities during a past period and determines whether the activities in the period are abnormal or not.
Formally, given a time stamp $t$ and the set of user activity sequences $\mathcal D_t$ where all activities were collected before the time stamp $t$, post-hoc ITD typically trains a model $\mathcal F_\text{PH}$ to find all the abnormal activities in $\mathcal{D}_t$.
}
\begin{figure*}[t!]
  \centering
  \includegraphics[width=0.95\textwidth]{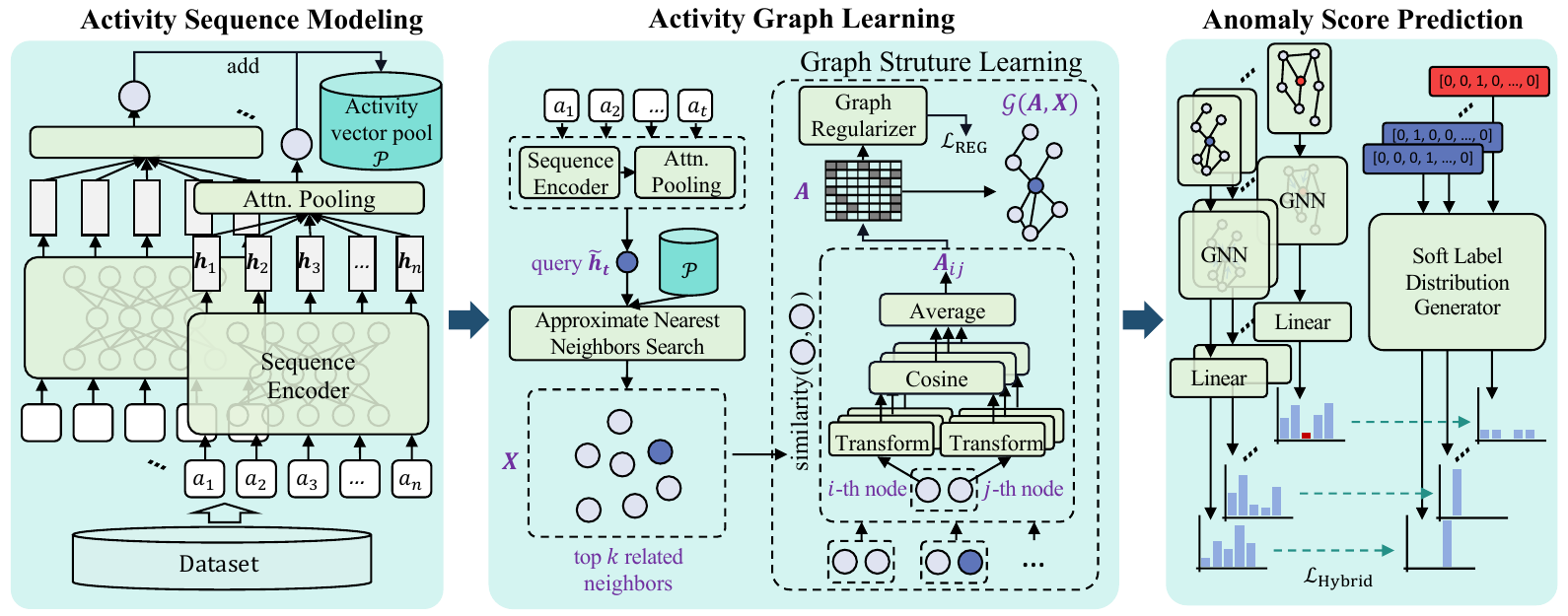}
  \caption{Overall architecture of \mname. }
  \label{fig:model_overview}
\end{figure*}
{
\underline{Real-Time Insider Threat Detection}. Real-time ITD aims at predicting whether the current activity of a user is abnormal based on previous activities in the system, which is usually deployed in real-time system monitoring. 
Formally, real-time ITD utilizes user activity sequence before the current time $t$, which is denoted by $\mathcal{D}_t$, to detect the abnormality of the current activity of each user.
Let $\mathcal A_{t}$ be the set of activities occurred at time $t$ of all users in the system, i.e., $\mathcal{A}_t=\{a^u \mid u\in \mathcal{U}, \text{time}(a^u) = t\}$, where $a^u$ represents one activity of user $u$, and $\text{time}(a^u)$ refers to the occurrence time of $a^u$.
Real-time ITD aims at developing an anomaly detection model $\mathcal{F}_\text{RT}$ based on $\mathcal{D}_{t}$ to discover abnormal activities in $\mathcal A_t$.
}

\section{Approach} \label{sec:app}
\subsection{Overview of \mname}
To ascertain the normalcy of the ongoing activity, we model the preceding sequence of activities and anticipate the current activity. A heightened probability in the prediction suggests that the current activity is less likely to be abnormal.
Formally, assuming user $u$ has generated an activity sequence of length $n$ as $\{a_1^u, a_2^u, \dots, a_n^u\}$, 
the goal of real-time ITD is to learn the activity patterns of normal users, model the probability of current activity $a_{n+1}^u$, and classify it as an insider threat if the probability falls below the detection threshold.

\Cref{fig:model_overview} depicts the overall architecture of \mname, which consists of {three modules, i.e., \moduleone, \moduletwo, and \modulethree.}
Learning the representation of a user's preceding activity sequence lies at the essence of real-time ITD.
In this paper, \mname first employs a sequence encoder to obtain preliminary activity representations.
However, utilizing only one sequence encoder can capture information for prediction solely from the user's historical activity sequence.
To enhance activity representations and reduce false positives, \mname queries the activity vector pool and introduces related (similar) activity vectors.
Then \mname constructs a graph among the activities and learns the graph structure adaptively.
Finally, to resolve the imbalance problem between normal and abnormal activities, we propose a hybrid prediction loss to incorporate supervised information from abnormal samples in addition to self-supervision with only normal samples.

The \mname \ architecture can be applied to both real-time ITD and post-hoc ITD.
However, these two scenarios involve distinct prediction paradigms while utilizing the same sequence encoder. Specifically, they employ next activity prediction for real-time ITD and activity cloze for post-hoc ITD.
We introduce the details of each module of \mname \ for real-time ITD from Section~\ref{sec:ie} to~\ref{sec:ad}
We explain the differences between post-hoc ITD and real-time ITD in Section~\ref{sec:ph}.

\subsection{\moduleone} \label{sec:ie}
In this section, we model the historical activity sequence of a user to learn the temporal dependencies among activities and obtain a preliminary representation of the sequence.
Specifically, we employ a sequential model for encoding activities and a multi-head attentive pooling layer to aggregate the information of the whole sequence.

\subsubsection{Sequence Encoder}
\label{Sec:Sequence Encoder}
Sequence models have achieved tremendous success in the field of natural language processing due to their strong ability to learn contextual representations. We use a sequence encoder to achieve a representation of the user's previous activities $S=\{a_1, \dots, a_{n}\}$.
We omit the superscript $u$ for brevity in the absence of ambiguity.
First, we assign a numeric token to each user activity according to its type and timestamp. Recall that the activity type could be \texttt{open file, connect device, login}, etc., as illustrated in \Cref{fig:fig1a_illustration}.
Following \cite{DBLP:conf/iscc/HuangZLLWY21}, we divide a day into 24 time slots by hour, and map the timestamp to an integer between 0 and 23. 
Formally, the activity code $c_i$ is obtained by:
\begin{equation}
    c_{i} = \text{type}(a_i)\times 24 + \text{time}(a_i) \,,
    \label{eq:code}
\end{equation}
where $\text{type}(a_{i})$ represents the activity type ID of $a_{i}$, and $\text{time}(a_{i})$ the corresponding time slot of $a_{i}$'s timestamp.
By converting $a_i$ to $c_i$ in this manner, we integrate the tasks of predicting activity type and occurrence time slot.

Then we project the activity code to an embedding space with an embedding layer. 
We initialize an embedding matrix $\bm W_E = [\bm w_1, \bm w_2, \dots, \bm w_M] \in \mathbb{R}^{d\times M}$, where $d$ is the size of embedding vectors, and $M$ the number of unique activities in the system.
For each activity code $c_i$, its embedding $\bm e_i$ is obtained by looking up the embedding matrix, i.e., 
\begin{equation}
    \bm e_i = \text{Embedding}(c_i) = \bm w_{c_i} \in \mathbb{R}^{d} \,,
\end{equation}
where $\bm e_i$ refers to the embedding vector of activity $a_i$.

To achieve activity representations, several sequence modeling methods can be employed, e.g, Long-Short Term Memory (LSTM)~\cite{hochreiter1997long}, Gated Recurrent Unit (GRU)~\cite{cho2014learning}, Transformer~\cite{vaswani2017attention}, etc.
Without loss of generality, we use LSTM for instance.
{The LSTM encodes the sequence of activity embeddings $\bm E = (\bm e_1, \bm e_2, \dots, \bm e_n)$ and generates a sequence of hidden states correspondingly:
\begin{equation}
\bm H = (\bm h_1, \bm h_2, \dots, \bm h_n) = \text{LSTM}(\bm e_1, \bm e_2, \dots, \bm e_n) \,.
\end{equation}}
We investigate the performance of different sequence encoders on \mname in Section~\ref{Sec:Compatibility Analysis}.

\subsubsection{Multi-Head Attentive Pooling}
\label{Sec:Multi-Head Attentive Pooling}
To enable the model to focus on different aspects of the past hidden states while capturing various dependency relationships between historical user activities, we employ multi-head attentive pooling to aggregate contextual information of past hidden states.

Given a query $\bm{q}$ and a set of keys $\bm{K}$ and values $\bm{V}$, where $\bm{q} \in \mathbb{R}^d$, $\bm{K}, \bm{V} \in \mathbb{R}^{n \times d}$, the scaled dot-product attention operation~\cite{vaswani2017attention} first computes attention scores as follows:
\begin{equation}
\alpha(\bm{q}, \bm{K}_i) = \frac{\bm{q}^\top \bm{K}_i}{\sqrt{d}} \,.
\end{equation}
Then it employs the normalized scores as weights to aggregate representations of preceding activities:
\begin{equation}
\text{Attention}(\bm{q}, \bm{K}, \bm{V}) = \sum_{i} \frac{\alpha(\bm{q}, \bm{K}_i)}{\sum_j \alpha(\bm{q}, \bm{K}_j)} \cdot \bm{V}_i\,.
\end{equation}

Multi-head attentive pooling enhances scaled dot-product attention by dividing it into multiple heads. Each head learns different feature representations and attention weights independently, focusing on different positions and semantics in the input sequence, boosting the model's expressive power:
\begin{equation}
\begin{aligned}
&\text{MHA}(S) = \text{Concat}(\text{head}_1, \text{head}_2, \dots, \text{head}_h) \cdot \bm{W}^O\,, \\
&\text{head}_i = \text{Attention}(\bm{q}_{\text{head}_i}, \bm{K}_{\text{head}_i}, \bm{V}_{\text{head}_i})\,, \\
&\bm{q}_{\text{head}_i} = \bm{q} \cdot \bm{W}^q_i, \quad \bm{K}_{\text{head}_i} = \bm{K} \cdot \bm{W}^K_i, \quad \bm{V}_{\text{head}_i} = \bm{V} \cdot \bm{W}^V_i\,,
\end{aligned}
\end{equation}
where $\bm{q}_{\text{head}_i} \in \mathbb{R}^{d_k}$, $\bm{K}_{\text{head}_i}, \bm{V}_{\text{head}_i} \in \mathbb{R}^{n \times d_k}$ 
, $d_k = \frac{d}{h}$ is the dimension after projection, $h$ is the number of heads, $\bm{W}^q_i, \bm{W}^K_i, \bm{W}^V_i \in \mathbb{R}^{d \times d_k}$ are the weight matrices for the linear projection of the $i$-th head, and $\bm{W}^O \in \mathbb{R}^{hd_k \times d}$.

To consolidate the semantics of a user's activities,
we consider the hidden state $\bm h_n$ at the last position of LSTM as the query $\bm{q}$ and all hidden states $\bm H = (\bm h_1, \bm h_2, \dots, \bm h_n)$ as the keys $\bm{K}$ and values $\bm{V}$. 
Following pooling with the multi-head attention, we derive the aggregated representation $\tilde{\bm{h}}_n$ for the $n$-th activity.

\subsection{\moduletwo} 
\label{Sec:module2}

Although the sequence-based methods can be applied to predict the next activity directly, they fail to consider the relationships among all activity sequences within the system.
Such relationships could enrich the semantics of the activity representations, alleviating false positives as a result.
To capture the relationships of activity sequences, we construct an activity graph specific to each detected activity.
Unfortunately, constructing a static graph among activity sequences is sub-optimal.
First, it is not suitable for real-time ITD, where user activities are constantly evolving.
Second, static graph construction requires expert knowledge, which may result in poor scalability and high cost.
Insight of this, we propose to automatically construct a dynamic graph that connects the detected activity with its closely associated activities, achieved through graph structure learning.

\subsubsection{Activity Vector Pool}
\label{Sec:Intention Vector Pool}

First, we apply the \moduleone \ module to all activity sequences within the system to achieve a pool of activity representation vectors.
Specifically, we split an activity sequence of a user into several sessions, as previous studies did \cite{DBLP:conf/iscc/HuangZLLWY21}.
To achieve the representation of each activity, we further split each activity sequence into several sub-sequences for real-time ITD.
For instance, given a session of user $u$ with a length of $l$, $S_{\text{session}}=\{a_1, a_2, \dots, a_l\}$, we obtain $l$ sub-sequences, where the $i$-th sub-sequence include the $i$-th activity and its preceding activities, i.e.,
$ \{a_{1}, a_{2}, \dots, a_{i}\} $.
By applying the \moduleone \ module to these sub-sequences, we can obtain an activity vector pool within the system, which is denoted by $\mathcal{P_{\text{RT}}}\in \mathbb{R}^{M\times d}$.
Recall that $M$ is the number of unique activities in the system, and $d$ is the size of the vectors.
The vectors in the vector pool $\mathcal{P_{\text{RT}}}$ are initialized randomly and optimized together with the remaining model.
{We include all the vectors of normal activities in the training set into the activity vector pool and filter out duplicate user activities.}

\subsubsection{Graph Structure Learning}
\label{Sec:Graph Structure Learning}
We retrieve the activity vector pool $\mathcal{P_{\text{RT}}}$ with $\tilde{\bm h}_n$ to obtain top $k$ most related vectors of the detected activity $a_n$, i.e, $\{\bm v_1, \bm v_2, \dots, \bm v_k\}, \bm v_i \in \mathbb{R}^{d}$.
To be specific, we use the approximate nearest neighbor searching algorithm to reduce retrieval time.
We then construct an activity graph $\mathcal{G}=(\bm A,\bm X)$ based on $\tilde{\bm h}_n$ and the top $k$ most related vectors,
where $\bm A \in \mathbb{R}^{(k+1)\times (k+1)}$ is the adjacency matrix of the graph, and $\bm X = [\tilde{\bm h}_n, \bm v_1, \bm v_2, \dots, \bm v_k] \in \mathbb{R}^{(k+1)\times d}$ is the node representations.

The matrix $\bm A$ is further optimized by the Graph Structure Learning component.
It learns a function $f_\mathcal{G}(\cdot,\cdot)$ that maps the connectivity relationship between any two nodes to a real-valued measurement.
For the $i$-th and $j$-th nodes with feature vectors $\bm x_i$ and $\bm x_j \in \bm X$, one simple measurement is to calculate the cosine similarity between the two vectors, i.e., $f_\mathcal{G}(i,j)=\cos(\bm x_i, \bm x_j)$.
We harness the concept of multi-head attention by employing a multi-head variant of weighted cosine similarity, expressed as:
\begin{equation}
\label{eq:weightedcosine}
f_\mathcal{G}(i,j) = \frac{1}{Z}\sum_{z=1}^Z\cos(\bm x_i \odot\bm{w}_{\text{GSL}}^z, \bm x_j \odot\bm{w}_{\text{GSL}}^z) \,,
\end{equation}
where $\odot$ represents the Hadamard product operation, $Z$ is the number of attention heads, $\cos(\cdot,\cdot)$ is the cosine similarity function, and $\bm W_{\text{GSL}}=[\bm w_{\text{GSL}}^1,\bm w_{\text{GSL}}^2, \dots,\bm w_{\text{GSL}}^Z]\in \mathbb R^{Z\times d}$ is the learnable weight matrix. The multi-head version of weighted cosine similarity allows the model to consider node relationships from different perspectives jointly.
To ensure each element of $\bm A$ to be non-negative, we filter the values of $f_\mathcal{G}(\cdot,\cdot)$ that are negative 
and set a hard threshold $\epsilon$ to suppress the noise from neighbors, i.e.,
\begin{equation}
\bm A_{ij} =
\begin{cases}
f_\mathcal{G}(i, j), &f_\mathcal{G}(i, j)\ge\epsilon \,,\\
0, &\text{otherwise} \,.
\end{cases}
\label{eq:aij}
\end{equation}

The graph structure is also learned together with other modules of \mname. 
It is capable of adapting to continuously occurring activities.
Additionally, compared to constructing a global graph, we only construct a local graph, but incorporate the most related activity sequences,
which is computationally efficient and practical.

\subsubsection{Graph Regularization}
\label{Sec:Graph Regularization}
Graph signals exhibit smooth variations between neighboring nodes \cite{DBLP:journals/pieee/OrtegaFKMV18}.
Following ~\cite{DBLP:journals/corr/Kalofolias16,DBLP:conf/nips/0022WZ20}, we employ Dirichlet energy~\cite{DBLP:conf/nips/BelkinN01} to regularize the smoothness of the activity graph $\mathcal G$.
A smaller Dirichlet energy indicates greater similarity and smoother graph signals, while a larger value indicates greater differences between adjacent nodes.
The regularizer is defined as follows:

\begin{equation}
\mathcal{L}_{D}=\frac{1}{2}\sum_{i \in \mathcal{V}}\sum_{j \in \mathcal{N}_i}\bm A_{ij} \left\lVert \bm{x}_i-\bm{x}_j \right\rVert^2 = \text{tr}(\bm{X}^\top\cdot\bm{L}\cdot\bm{X}) \,,
\end{equation}
where $\mathcal{V}$ denotes the set of vertices in the graph $\mathcal{G}$, $\mathcal{N}_i$ represents the neighborhood of node $i$, $\bm{A}_{ij}$ represents the connectivity between nodes $i$ and $j$ in the graph, 
$\bm x_i$ and $\bm x_j$ are the node representations, $\text{tr}(\cdot)$ represents the trace of a matrix, and $\bm{L} = \bm{D} - \bm{A}$ represents the Laplacian matrix of the graph, where $\bm{D}$ is the degree matrix.
Besides, we can use $\hat{\bm{L}} = \bm{D}^{-\frac{1}{2}} \bm{L} \bm{D}^{-\frac{1}{2}}$ instead of $\bm{L}$ to make the smoothness invariant to node degrees \cite{chung1997spectral}.

Minimizing the Dirichlet energy penalizes the degree of connectivity between dissimilar nodes and encourages graphs with smooth signals to correspond to a sparse set of edges. In extreme cases, this can lead to a trivial solution, i.e., $\bm{A}=\bm{0}$.
To ensure meaningful learned graphs, we impose a constraint on graph connectivity.
Following \cite{DBLP:journals/corr/Kalofolias16}, we add a logarithmic barrier term in the graph regularization loss:
\begin{equation}
\mathcal{L}_{log}=-\bm 1^\top \cdot\log(\bm A\cdot\bm 1)\,.
\end{equation}
Additionally, to directly control sparsity, we follow \cite{DBLP:conf/nips/0022WZ20} and append the Frobenius norm:
\begin{equation}
\mathcal{L}_{sparsity}=\left\lVert \bm{A} \right\rVert^2_F\,.
\end{equation}

Finally, the whole regularization loss is as follows:
\begin{equation}
\begin{aligned}
\mathcal{L}_{\text{Reg}}=& \frac{\mu_1}{n^2}\mathcal{L}_D + \frac{\mu_2}{n} \mathcal{L}_{log} + \frac{\mu_3}{n^2}\mathcal{L}_{sparsity} \\
=&\frac{\mu_1}{n^2}\text{tr}(\bm{X}^\top\cdot\bm{L}\cdot\bm{X}) - \frac{\mu_2}{n}\bm 1^\top \cdot\log(\bm A\cdot\bm 1)+ \frac{\mu_3}{n^2} \left\lVert \bm{A} \right\rVert^2_F \,.
\end{aligned}
\end{equation}
where $\mu_1, \mu_2, \mu_3$ are the hyperparameters.

\subsection{\modulethree} \label{sec:ad}

To detect abnormal activities, we leverage graph neural networks (GNN) to enrich the activity representation from the activity graph.
Then we build a fully connected neural network to predict the activity code ($c_i$ in Equation~\eqref{eq:code}) after the GNN.
Additionally, we propose a novel hybrid prediction loss function to resolve the significantly imbalanced classes in real-time ITD.

\subsubsection{Graph Neural Network}
\label{Sec:Graph Neural Network}
We utilize a GNN model to learn node embeddings in the activity graph.
The GNN takes the activity graph $\mathcal{G}$ as input and applies a message passing mechanism to capture the dependencies between the nodes. 
The generalized GNN can be seen as a stack of layers composed of Aggregation steps and Update steps:
\begin{equation}
\begin{aligned}
\bm{n}_{i}^{(p)}&=\underset{j \in \mathcal{N}_{i}}{\text {Aggregator}_{p}} \left(\bm{x}_{j}^{(p)}\right) \,, \\
\bm{x}_{i}^{(p+1)}&=\operatorname{Updater}_{p}\left(\bm{x}_{i}^{(p)}, \bm{n}_{i}^{(p)}\right) \,,
\end{aligned}
\end{equation}
where $\text{Aggregator}_p$ and $\text{Updater}_p$ represent the Aggregation and Update operations at the $p$-th layer, $\bm{x}_i^{(p)}$ represents the representation of node $i$ at the $p$-th layer, $\mathcal{N}_i$ is the neighborhood of node $i$, and $\bm{n}_i^{(p)}\in \mathbb{R}^{d}$ refers to the aggregated information from the neighbors of node $i$ at the $p$-th layer. 

In this paper, we explore two widely-used GNN architectures to optimize node embeddings, i.e., Graph Convolutional Network (GCN)~\cite{kipf2016semi} and Graph Attention Network (GAT)~\cite{velivckovic2018graph}. For GCN, the Aggregation step and Update step are formulated as follows: 
\begin{equation}
\begin{aligned}
\bm{n}_{i}^{(p)}&=\sum_{j \in \mathcal{N}_{i}} \bm D_{i i}^{-\frac{1}{2}} \bm A_{ij}\bm D_{j j}^{-\frac{1}{2}} \bm{x}_{j}^{(p)} \,, \\
\bm{x}_{i}^{(p+1)}&=\delta\left({\bm{W}^{(p)}_{\text{GCN}}} \bm{n}_{i}^{(p)}\right) \,,
\end{aligned}
\end{equation}
where $\bm{A}_{ij}$ represents the edge weight between nodes $i$ and $j$ in the graph, $\bm{D}_{ii} = \sum_{j=1}^{k+1} \bm{A}_{ij}$, $\bm{W}^{(p)}_{\text{GCN}}\in \mathbb{R}^{d\times d}$ is a weight matrix at the $p$-th layer, and $\delta(\cdot)$ represents a non-linear activation function, e.g, Rectified Linear Unit (ReLU)~\cite{nair2010rectified}.

The Aggregation step and Update step of GAT are formulated as follows:
\begin{gather}
\bm{n}_{i}^{(p)}=\sum_{j \in \mathcal{N}_{i}} \gamma_{i j} \bm{x}_{j}^{(p)} \,, \notag \\
\gamma_{ij}=\frac{\exp \left(\beta\left({\bm C^{(p)}}^\top \left[{\bm{W}^{(p)}_{\text{GAT}}} \bm{x}_{i}^{(p)}; {\bm{W}^{(p)}_{\text{GAT}}} \bm{x}_{j}^{(p)}\right]\right)\right)}
{\sum_{j^\prime \in \mathcal{N}_{i}} \exp \left(\beta\left({\bm C^{(p)}}^\top \left[{\bm{W}^{(p)}_{\text{GAT}}} \bm{x}_{i}^{(p)}; {\bm{W}^{(p)}_{\text{GAT}}} \bm{x}_{j^\prime}^{(p)}\right]\right)\right)} \,, \\
\bm{x}_{i}^{(p+1)}=\delta\left({\bm{W}^{(p)}_{\text{GAT}}} \bm{n}_{i}^{(p)}\right) \notag \,,
\end{gather}
where $\gamma_{ij} \in \mathbb{R}$ represents the importance of node $j$ to node $i$ (i.e., the attention weight), $\bm{C}^{(p)} \in \mathbb{R}^{2d}$ is a weight vector of a linear layer, $\beta(\cdot)$ stands for the Leaky Rectified Linear Unit (Leaky ReLU) activation function~\cite{maas2013rectifier}, $\bm{W}_{\text{GAT}}^{(l)} \in \mathbb{R}^{d\times d}$ is a shared learnable weight matrix at the $p$-th layer used to provide sufficient expressive power,
and $\delta(\cdot)$ refers to the ReLU activation function.

After applying GCN or GAT to the activity graph $\mathcal{G}=(\bm A, \bm X)$, the ongoing activity vector $\tilde{\bm h}_n$ further aggregates information from the top $k$ most related neighbors.
We obtain an enhanced vector after the graph neural network, which is denoted as $\tilde{\bm h}_n^\prime \in \mathbb R^{d}$.
Then, we use a fully connected layer to predict the next activity of the user, which is formulated as: 

\begin{equation}
\hat {\bm y}_{n+1} = \text{softmax}(\bm W_{\text{FC}}\bm \tilde{\bm h}_n^\prime + \bm b_{\text{FC}}) \,,
\end{equation}
where $\bm{W}_{\text{FC}} \in \mathbb{R}^{M\times d}$ is the weight matrix, $\bm b_{\text{FC}} \in \mathbb{R}^{M}$ is the bias, and $\hat {\bm y}_{n+1} \in \mathbb{R}^{M}$ represents the probability distribution of the predicted activity.
We compare the probability corresponding to the current activity with the detection threshold to determine if it is anomalous.

\subsubsection{Hybrid Prediction Loss}
\label{Sec:Hybrid Prediction Loss}

In a real enterprise system, abnormal activities are very limited and difficult to identify,
leading to significantly imbalanced sample numbers between normal activities and abnormal ones~\cite{ding2023tmg}.
Therefore, alleviating the impact of data imbalance is crucial for ITD.

In this paper, we alleviate the impact of data imbalance by leveraging the supervision of the limited abnormal activity labels. 
Given the substantial class imbalance, it is inappropriate to treat the ITD problem as a binary classification task due to the inadequacy of abnormal labels.
The LAN paradigm leverages historical activities to predict the next one, deeming an activity abnormal if its occurrence probability is exceedingly low. This approach seamlessly integrates self-supervised signals derived from normal labels, fostering natural learning of normal activity patterns. 
However, training the model in this self-supervised manner introduces noise if the abnormal activities are treated as normal ones.
Therefore, to improve the performance of ITD, we propose a novel hybrid prediction loss that combines both supervised and self-supervised learning,
imparting awareness to the self-supervised model regarding labels associated with abnormal activities.

Specifically, given an activity sequence $S = \{a_1, a_2, \dots, a_n\} $ and the corresponding anomaly labels of the activities $\bm q \in \mathbb{R}^{n}$, $\bm q_i \in \{0, 1\}$, the conventional approach for optimizing the next activity prediction is to use one-hot encoding of activities and the cross-entropy loss, which is defined by $\mathcal{L}_{\text{CE}} = -\frac{1}{n-1} \sum_{i=2}^{n} \log(\hat{\bm Y}_{i,c_i})$,
where $\hat{\bm Y}\in \mathbb{R}^{n \times M}$ represents the probability distribution of the model's predictions for the occurrence of behaviors at each position. $\hat{\bm Y}_i \in \mathbb{R}^{M}$ corresponds to the i-th position, and $\hat{\bm Y}_{i,c_i}$ denotes the predicted probability of the occurrence of the ground truth activity $a_{i}$ (the $c_i$-th item in $\hat{\bm Y}_i$). Recall that $c_i$ is the corresponding integer code of $a_{i}$.

However, this cross-entropy loss only optimizes the positive feedback of predicting the next activity.
To explicitly utilize the negative feedback provided by anomaly labels, we take the labels of abnormal activities (i.e., $q_{i}=1$) into account. 
Specifically, for abnormal user activities, we do not want the model to learn similar activity patterns.
Thus, we prefer to suppress the occurrence probability of such anomalous activities compared to other ones.
We achieve this goal indirectly by increasing the probabilities of other activities under the constraint of softmax, thus reducing the occurrence probability of anomalous activities.

We construct a new soft label distribution, which can be calculated by:
\begin{equation}
\bm Y^\prime = \bm Y\odot(1-\bm \Omega)+\frac{1}{M-1}\bm \Omega\odot(1-\bm Y\odot\bm \Omega) \,,
\label{eq:soft_label}
\end{equation}
where $\bm{\Omega} = \bm{q} \otimes \bm{1}_M^\top \in \mathbb{R}^{n \times M} $ is a masking matrix, $\otimes$ represents the Kronecker product operation, $\bm{1}_M \in \mathbb{R}^{M}$ denotes a $M$-dimensional vector filled with ones.
$\bm{Y} \in \mathbb{R}^{n\times M}$ represents the one-hot labels of activity sequence $S$. When $\bm q_{i}=1$, the resulting label distribution from Equation~\eqref{eq:soft_label} can be simplified as:
\begin{equation}
\bm Y_{ij}^\prime = \begin{cases}              0, &              j=c_i\,, \\                \frac{1}{M-1},  &                j\neq c_i \,.        \end{cases}
\end{equation}
Furthermore, to ensure that the learning of normal activities is not affected, we keep the probability distribution of normal activities unchanged, i.e., $\bm Y_{ij}^\prime  = \bm Y_{ij}$, when $\bm q_{i}=0$.

Furthermore, the importance of abnormal activity samples is different from that of normal samples. We assign a weight $r$ to the abnormal samples and construct a sample weight vector $\bm{w}^{s}$, where $\bm{w}^{s}_{i} = 1+(r-1) \cdot \bm q_{i}$, that is:
\begin{equation}
\label{eq:hardweight}
\bm w^{s}_{i}= \begin{cases}              1, &              \bm q_{i}=0 \,, \\                r,  &                \bm q_{i}=1   \,,       \end{cases}
\end{equation}
We refer to this operation as the weighting negative feedback operation for the next activity prediction task.

Finally, the hybrid prediction loss is defined as a weighted soft cross-entropy, i.e.,
\begin{equation}
\mathcal{L}_{\text{Hybrd}} = -\frac{1}{n-1} \sum_{i=2}^{n} \bm w^{s}_{i}\sum_{j=1}^{M} \bm Y_{ij}^\prime \log(\hat{\bm Y}_{ij}) \,.
\end{equation}

The overall loss function of \mname \ is:
\begin{equation}
\mathcal{L}=  \mathcal{L}_{\text{Hybrid}}+ \mathcal{L}_{\text{Reg}} \,,
\end{equation}

We optimize the model parameters by minimizing $\mathcal{L}$.

\subsection{Post-Hoc ITD} 
\label{sec:ph}
In real-time ITD, we utilize the user's previous activity sequence to predict the next activity. However, in post-hoc ITD, we need to make some changes. Instead of predicting the next activity, we treat it as an ``activity cloze'' task. For a certain time step $t$, we mask the user's activity at that time step, resulting in a new activity sequence $S'=\{a_1, \dots, a_{t-1}, \langle\textit{MASK}\rangle, a_{t+1}, \dots, a_n\}$. We feed this new $S'$ into a sequence encoder with bidirectional contextual capabilities while keeping the rest of the \mname \  unchanged. Then, we predict the masked activity to complete the detection.

\section{Experiments} \label{sec:ex}

\subsection{Experimental Setup}

\subsubsection{Dataset}
As previous studies~\cite{ DBLP:conf/ccs/LiuWZJXM19,DBLP:journals/tifs/LiLJLYGY23,yuan2020few, alslaiman2023enhancing,le2020analyzing,yuan2021deep}, we evaluate the performance of \mname on two publicly available datasets, i.e., CERT r4.2 and CERT r5.2~\cite{Lindauer2020}.
With different scales, CERT r4.2 and CERT 5.2 both contain user activity data in a company from January 2010 to June 2011. \Cref{tab:statistics} summarizes the statistics of the datasets. Specifically, CERT r4.2 contains 1,000 employees with 32,770,222 user activities, among which 7,316 activities of 70 employees were manually injected as abnormal activities by domain experts.  
Similarly, CERT r5.2 contains 2,000 employees with 79,856,664 activities, and 10,306 activities of 99 employees were manually injected as abnormal activities.
It can be seen that normal employees and normal activities occupy the vast majority of the whole dataset. To clearly show the data imbalance problem, we calculate the imbalance ratio (IR) by $N_{maj}/N_{min}$, where $N_{maj}$ and $N_{min}$ are the sample sizes of the majority class and the minority class, respectively~\cite{buda2018systematic}. The larger
the IR, the greater the imbalance extent of the dataset. 
From \Cref{tab:statistics} it can be seen that the IR is 13 for employees in CERT 4.2. When it comes to the activities, the IR becomes 4,478 in the same dataset, which indicates the difficulty in classifying normal and abnormal activities. For CERT r5.2, the data imbalance is even worse, which poses great challenges for activity-level ITD.  

\subsubsection{Data Preprocessing}
Since there are multiple sources for user activity logs (e.g., logs of logon or website visit), the first step of data preprocessing is to aggregate the data from all sources so that the activities of the same user can be aggregated according to their time stamps. 
Then, as previous studies~\cite{yuan2019insider,vinay2022contrastive}, we split the activities of each user into sessions, where each session contains a set of activities in chronological sequence between a user's logon and logoff. Since the goal is to detect insider threats at the activity level, each activity is regarded as a sample for training the model. Specifically, for real-time ITD, each activity is represented by the activity itself along with its previous activities in the same session, so that an abnormal activity can be immediately detected once it occurs. 
As for post-hoc ITD, we put each activity into its corresponding session, and an activity can only be detected after the session ends.
Moreover, we filter out repetitive HTTP operations within the same hour following~\cite{DBLP:conf/iscc/HuangZLLWY21}. The statistics for data after preprocessing can also be seen in~\Cref{tab:statistics}. 
Finally, we use the user activity data in 2010 for training and validation, and use the data from January 2011 to June 2011 for evaluating all methods.

\begin{table}
\footnotesize
    \centering
    \caption{Statistics of the datasets}
    \label{tab:statistics}
    \begin{tabular}{lrr}
        \toprule
        Dataset & CERT r4.2 & CERT r5.2  \\
        \midrule
         \# Normal Employees & 930 & 1,901\\ 
         \# Abnormal Employees & 70 & 99\\ 
         \rowcolor{gray!20} 
          \begin{tabular}[c]{@{}c@{}}\textbf{Imbalance Ratio}\end{tabular} & \textbf{13} & \textbf{19}\\ \hline 
         \# Normal Activities & 32,762,906 & 79,846,358 \\
         \# Abnormal Activities & 7,316 & 10,306 \\
         \rowcolor{gray!20} 
         \begin{tabular}[c]{@{}c@{}}\textbf{Imbalance Ratio}\end{tabular} & \textbf{4,478} & \textbf{7,748} \\ \hline
         \# Normal Activities after Preprocessing&  7,664,484 & 27,254,280 \\
         \# Abnormal Activities after Preprocessing&  7,316 & 10,306 \\
         \rowcolor{gray!20} 
         \begin{tabular}[c]{@{}c@{}}\textbf{Imbalance Ratio} \end{tabular} & \textbf{1,048} &\textbf{2,645} \\
        \bottomrule
    \end{tabular}
\end{table}

\subsubsection{Baselines}
{To evaluate the performance of \mname on activity-level ITD, we compare \mname with competitive baselines that have the ability to detect insider threats at the activity level.} In this paper, we configure \mname as using LSTM as
the sequence encoder and using GCN as the GNN.
For real-time ITD, we compare \mname with 9 state-of-the-art methods, i.e., RNN~\cite{elman1990finding}, GRU~\cite{chung2014empirical}, DeepLog~\cite{du2017deeplog} (LSTM~\cite{hochreiter1997long}), Transformer~\cite{vaswani2017attention}, RWKV~\cite{peng2023rwkv}, TIRESIAS~\cite{shen2018tiresias}, DIEN~\cite{zhou2019deep}, BST~\cite{chen2019behavior}, and FMLP~\cite{zhou2022filter}. 
{Specifically, DeepLog utilizes LSTM to predict whether each log entry is anomalous. RNN, GRU, LSTM, and Transformer are four widely used architectures for sequence modeling. 
RWKV is a recently proposed sequence modeling architecture that combines the advantages of efficient parallel training in Transformer and efficient sequential inference in RNN. 
DIEN and BST are commonly used methods for user behavior modeling, which have been widely used for click-through rate prediction in recommendation systems.
FMLP, a sequential recommendation model, filters out noise from user historical activity data to reduce model overfitting. 
}

For post-hoc ITD, we compare \mname with 8 state-of-the-art methods, i.e., RNN, GRU, DeepLog(LSTM), Transformer, FMLP, ITDBERT~\cite{DBLP:conf/iscc/HuangZLLWY21}, OC4Seq~\cite{wang2021multi}, and log2vec~\cite{DBLP:conf/ccs/LiuWZJXM19}. {Unlike real-time ITD, for the post-hoc ITD task we configure RNN, GRU, and DeepLog as bidirectional models to capture both preceding and succeeding contexts of each activity simultaneously. ITDBERT uses an attention-based LSTM for session-level prediction, using attention weights of each activity as the anomaly score. 
OC4Seq regards log anomaly detection as a one-class classification problem. It represents user activities using RNN and trains the model only on normal activities, searching for an optimal hypersphere in the latent space to enclose normal activities. During inference, it predicts whether an activity falls within the hypersphere to determine if it is anomalous.
Log2vec employs graph-based methods to detect malicious activities. It designs heuristic rules to manually extract edges to represent relationships between activities. 
We reproduce log2vec following the instructions in ~\cite{DBLP:conf/ccs/LiuWZJXM19}.}

\subsubsection{Evaluation Metrics}
Like previous studies{~\cite{buczak2015survey,le2020analyzing,le2021anomaly,DBLP:conf/ndss/KingH22}}, we utilize the Receiver Operating Characteristic (ROC) curves to visualize the detection rate (DR) and the false positive rate (FPR) of each model, where $DR=TP/(TP+FN)$ and $FPR=FP/(FP+FN)$. TP, FN, FP, and TN represent the number of true positives, false negatives, false positives, and true negatives, respectively.
We compute the Area Under Curve (AUC) to evaluate the overall performance of all methods. 
Note that for practical use, it is necessary to fix a decision threshold for anomaly detection, i.e., activities with lower likelihood scores than the decision threshold are regarded as abnormalities. To achieve a trade-off between DR and FPR, a common practice is to set the decision threshold using the Youden index~\cite{youden1950index}, which can be calculated by $DR-FPR$.
We use the decision threshold with the maximum Youden index and report the DR and FPR according to the threshold.
{In addition, following previous studies~\cite{le2021anomaly,le2021exploring,tuor2017deep}, we also vary the investigation budget, which is the amount of suspicious activities that security analysts can inspect, and set the decision threshold accordingly.
Specifically, we vary the investigation budget by 5\%, 10\%, and 15\% of the number of activities to be tested, and report the corresponding detection rates DR@5\%, DR@10\%, and DR@15\%, respectively.}

\begin{figure*}[h]
  \centering
    \includegraphics[width=1\textwidth]{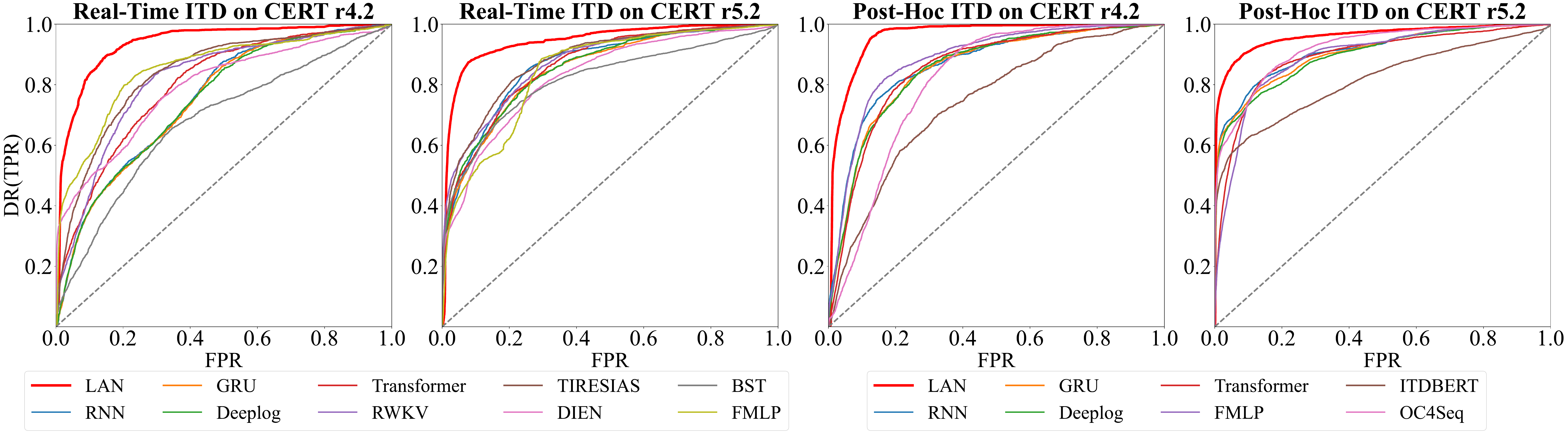}
  \caption{ROC curves for real-time ITD and post-hoc ITD on two datasets}
  \label{fig:ROC}
\end{figure*}

\begin{table*}[h!]
    \centering
    \caption{Performance comparison of LAN with nine baselines for real-time ITD. The best and second-best results are boldfaced and underlined, respectively. An upward arrow indicates the higher the better, and a downward arrow indicates the lower the better.}
    \label{tab:Real-time-Main}
    \begin{adjustbox}{width=1\textwidth}
    \setlength{\tabcolsep}{2pt}
    \begin{tabular}{cc@{\hspace{14pt}}c@{\hspace{14pt}}c@{\hspace{10pt}}ccc|c@{\hspace{14pt}}c@{\hspace{14pt}}c@{\hspace{10pt}}ccc}
        \toprule
        \multirow{2}{*}{\textbf{Model}} & \multicolumn{6}{c|}{\textbf{CERT r4.2}} & \multicolumn{6}{c}{\textbf{CERT r5.2}}   \\
        \cmidrule{2-13}
         & \textbf{AUC} $\uparrow$ & \textbf{DR} $\uparrow$ & \textbf{FPR} $\downarrow$ & \textbf{DR@5\%}$\uparrow$ & \textbf{DR@10\%}$\uparrow$ & \textbf{DR@15\%}$\uparrow$ & \textbf{AUC} $\uparrow$ & \textbf{DR} $\uparrow$ & \textbf{FPR} $\downarrow$  & \textbf{DR@5\%}$\uparrow$ & \textbf{DR@10\%}$\uparrow$ & \textbf{DR@15\%}$\uparrow$\\
        \midrule
        RNN\cite{elman1990finding}  & 0.7521 & 0.6934 & 0.3622& 0.2299 & 0.3821 & 0.4625 & 0.8641 & 0.8286 & 0.2361 & 0.4548 & 0.5928 & 0.6910\\
        GRU\cite{chung2014empirical}  & 0.7486 & 0.7119 & 0.3804 & 0.2391 & 0.3815 & 0.4614 &  0.8504 & 0.7911 &  0.2395 & 0.4637 & 0.5704 &0.6499\\
        DeepLog\cite{du2017deeplog} & 0.7469 & 0.7152 & 0.3767 & 0.2310 & 0.3842 & 0.4620 & 0.8549 & 0.7767 & 0.2336 & 0.4970 & 0.5954 & 0.6648\\
        Transformer\cite{vaswani2017attention} & 0.7981 & 0.7195 & 0.2799 & 0.2918 & 0.4201 & 0.5321 &  0.8628 & 0.7621& \underline{0.1985} & 0.4858 & 0.5745 & 0.6694\\
        RWKV\cite{peng2023rwkv} & 0.8165 & 0.7923 & 0.2576 & 0.2630 & 0.4348 & 0.5886 & 0.8727 & 0.8020 & 0.2345 & 0.5380 & 0.6224 & 0.6887\\
        TIRESIAS\cite{shen2018tiresias} & \underline{0.8377} & 0.7820 & 0.2338 & 0.3761 & 0.5277 & 0.6484 & \underline{0.8804} & 0.8129 & 0.2073 & \underline{0.5463} & \underline{0.6373} & \underline{0.7297}\\
        DIEN\cite{zhou2019deep} & 0.7894 & 0.7461 & 0.3072  & 0.4147 & 0.4875 & 0.5342 & 0.8268 & 0.7724 & 0.2690 &  0.3811 & 0.5455 & 0.6175\\
        BST\cite{chen2019behavior} & 0.6777 & 0.6554 & 0.3451 & 0.1625 & 0.2614 & 0.3647 & 0.8162 & 0.7417 & 0.2301 & 0.4772 & 0.5650 & 0.6548\\
        FMLP\cite{zhou2022filter} & 0.8526 & \underline{0.7983} & \underline{0.2027} & \underline{0.4783} & \underline{0.5647} & \underline{0.6837} & 0.8435 & \underline{0.8757} & 0.2889 & 0.4278 & 0.5171 & 0.5659\\
        \textbf{\mname}(Ours) & \textbf{0.9369} & \textbf{0.8875} & \textbf{0.1411} & \textbf{0.6832} & \textbf{0.8337} & \textbf{0.8951} & \textbf{0.9439} & \textbf{0.8814} & \textbf{0.0867} & \textbf{0.8089} & \textbf{0.8881} & \textbf{0.9076}\\
    
        \midrule
        Abs. Improv. & 0.0992 & 0.0892 & 0.0616 & 0.2049 & 0.2690 & 0.2114 & 0.0635 & 0.0057 & 0.1118 & 0.2626 & 0.2508 & 0.1779\\
        Rel. Improv.(\%) & 11.84\% & 11.17\% & 30.39\% & 42.84\% & 47.64\% & 30.92\%& 7.21\% & 0.65\% & 56.32\% & 48.07\% & 39.35\% & 24.38\%\\
        \bottomrule
    \end{tabular}
    \end{adjustbox}
   \end{table*}

\subsubsection{Implementation Details}
To conduct a fair comparison, we set the same hidden layer size (i.e., 128) for all methods, and all models were trained on the same training set with the early stopping strategy for 10 epochs.
We select the best learning rate and use the AdamW optimizer~\cite{loshchilov2017decoupled} with a weight decay of 0.01 when training each model.
For the hyper-parameters in \mname, we use 8 attention heads for multi-head attentive pooling, and the sizes of the activity vector pool are {1,025,920} and {5,359,987} for CERT r4.2 and CERT r5.2, respectively.
To efficiently retrieve the top $k$ vectors ($k$ is set to 15 after parameter analysis) most relevant to the current activity vector, we exploit Faiss~\cite{johnson2019billion}, a high-performance open-source library for searching approximate nearest neighbors in dense vector spaces, and utilize the Hierarchical Navigable Small Word (HNSW) algorithm~\cite{malkov2018efficient} to build the index, which has a logarithmic retrieval time complexity.
During graph structure learning, we utilize 4 attention heads in ~\Cref{eq:weightedcosine} and set the hard threshold $\epsilon$ to suppress noise in~\Cref{eq:aij} to 0.5 based on experimental experience.
As previous studies on graph structure learning~\cite{DBLP:conf/nips/0022WZ20}, the hyper-parameters $\mu_1$, $\mu_2$, and $\mu_3$ for the graph regularization loss are set to 0.2, 0.1 and 0.1, respectively.
For the weight $r$ in the hybrid prediction loss, we set it to the imbalance ratio of the dataset according to parameter analysis.
Since the weight $r$ is very large, to ensure numerical stability, we replicate the abnormal activity samples $r$ times to achieve the same effect as the weighted loss.

\subsection{Experimental Results}

\Cref{fig:ROC} depicts the ROC curves of all methods for real-time and post-hoc ITD. Overall, it can be seen that the ROC curves of \mname are much closer to the top-left corner than the curves of all the baselines in all experiments. The results indicate that \mname achieves better overall performance than all the baselines for both real-time and post-hoc ITD.

\Cref{tab:Real-time-Main} shows the detection results of \mname and 9 state-of-the-art baselines for real-time ITD. We use AUC, DR, FPR, DR@5\%, DR@10\%, and DR@15\% to evaluate the performance of all methods. Among these metrics, the higher AUC, the better overall performance. For the detection metrics DR, DR@5\%, DR@10\%, and DR@15\%, a higher detection rate indicates that the model can identify more anomalies.
Finally, the lower FPR, the fewer false positives that waste the investigation budget.
It can be seen that \mname is superior to all the baselines with regard to all six metrics on two datasets, with the highest AUC and detection rates and the lowest FPRs among all methods. Specifically, \mname surpasses the baselines by at least 9.92\% and 6.35\% in AUC on CERT r4.2 and r5.2, respectively. For the FPR scores, it can be seen that the FPRs of \mname are 14.11\% and 8.67\% on CERT r4.2 and r5.2, respectively, lower than all the baselines by at least 6.16\% and 11.18\% on two datasets. The results indicate that \mname significantly reduces the number of false positives, which has great value when the investigation budget is finite. 
Finally, comparing \mname with DeepLog, it can be observed that although both of them utilize LSTM as the backbone, \mname achieves better performance by autonomously learning the global relationships between activities in different sequences. 

In addition to real-time ITD, we also apply \mname, with slight modifications, for post-hoc ITD. To evaluate the effectiveness of \mname for post-hoc ITD, we compare \mname with 8 state-of-the-art baselines (i.e.,  RNN~\cite{elman1990finding}, GRU~\cite{chung2014empirical}, DeepLog~\cite{du2017deeplog}, Transformer~\cite{vaswani2017attention}, FMLP~\cite{zhou2022filter},  ITDBERT~\cite{DBLP:conf/iscc/HuangZLLWY21},  OC4Seq~\cite{wang2021multi},  and log2vec~\cite{DBLP:conf/ccs/LiuWZJXM19}. 
\Cref{tab:Post-Hoc-Main} displays the results of all methods for post-hoc ITD on CERT r4.2 and r5.2. 
{Note that log2vec constructs a graph for each abnormal user and performs clustering for the log entries of each user, assuming that smaller clusters tend to be suspicious. For this reason, the detection of abnormal activities relies on how to determine the threshold that represents the size of clusters. As a result, log2vec can not be applied when the investigation budget is fixed (e.g., 5\% of all activities). Hence, we did not compute the DR@5\%, DR@10\%, and DR@15\% for log2vec.}

\begin{table*}[ht!]
    \centering 
    \caption{Performance comparison of LAN with eight baselines for post-hoc ITD. The best and second-best results are boldfaced and underlined, respectively. An upward arrow indicates the higher the better, and a downward arrow indicates the lower the better.}
    \label{tab:Post-Hoc-Main}    
    \begin{adjustbox}{width=1\textwidth}
    \setlength{\tabcolsep}{2pt}
    \begin{tabular}{cc@{\hspace{14pt}}c@{\hspace{14pt}}c@{\hspace{10pt}}ccc|c@{\hspace{14pt}}c@{\hspace{14pt}}c@{\hspace{10pt}}ccc}
    \toprule
     \multirow{2}{*}{\textbf{Model}} & \multicolumn{6}{c|}{\textbf{CERT r4.2}} & \multicolumn{6}{c}{\textbf{CERT r5.2}}  \\
    \cmidrule{2-13}
     & \textbf{AUC} $\uparrow$ & \textbf{DR} $\uparrow$ & \textbf{FPR} $\downarrow$ & \textbf{DR@5\%}$\uparrow$ & \textbf{DR@10\%} & \textbf{DR@15\%}$\uparrow$ & \textbf{AUC} $\uparrow$ & \textbf{DR} $\uparrow$ & \textbf{FPR} $\downarrow$ & \textbf{DR@5\%}$\uparrow$ & \textbf{DR@10\%} & \textbf{DR@15\%}$\uparrow$ \\
    \midrule
    RNN\cite{elman1990finding} & 0.8652 & 0.8032 & 0.1996  & \underline{0.4375} & 0.6707 & 0.7549 & 0.9108 & 0.8131 & \underline{0.1359} & \underline{0.6881} & \underline{0.7699} & 0.8230 \\
    GRU\cite{chung2014empirical} & 0.8514 & 0.8103 & 0.2378 & 0.3364 & 0.5962 & 0.7001 & 0.9040 & 0.7879 & 0.1367 & 0.6780 & 0.7458 & 0.7957\\
    DeepLog\cite{du2017deeplog}& 0.8531 & 0.7891 & 0.2259 & 0.3310 & 0.5908 & 0.6897 & 0.8985 & 0.7730 & 0.1385 & 0.6729 & 0.7329 & 0.7779\\
    Transformer\cite{vaswani2017attention} & 0.8533 & 0.8005 & 0.2112 & 0.3109 & 0.5451 & 0.6951 & 0.8929 & 0.8358  & 0.1586 & 0.5684 & 0.7357 & \underline{0.8247}\\
    FMLP\cite{zhou2022filter} &  \underline{0.8837} & \underline{0.8190} & \underline{0.1671} & 0.4266 & \underline{0.6772} & \underline{0.7957} & 0.8920 & 0.8169 & 0.1614 & 0.4878 & 0.7412 & 0.8066\\
    ITDBERT\cite{DBLP:conf/iscc/HuangZLLWY21}  & 0.7413 & 0.6911 & 0.3153 & 0.2005 & 0.3272 & 0.4383 & 0.8139 & 0.6853 & 0.1996 & 0.5724 & 0.6243 & 0.6518\\
    OC4Seq\cite{wang2021multi} & 0.8113 & 0.8080 & 0.2940  & 0.1466 & 0.3019 & 0.4712 & \underline{0.9202} & \underline{0.8503} & 0.1727 & 0.6414 & 0.7383 & 0.8245\\
    log2vec\cite{DBLP:conf/ccs/LiuWZJXM19} & 0.6563  & 0.6793 & 0.3022 & - & -& - &  0.6178 & 0.6441 & 0.3388  & - & - & -\\
    \textbf{\mname}(Ours) & \textbf{0.9607} & \textbf{0.9478} & \textbf{0.1222} & \textbf{0.7429} & \textbf{0.8929 } & \textbf{0.9739} & \textbf{0.9605} & \textbf{0.9024} & \textbf{0.0865} & \textbf{0.8591} & \textbf{0.9088} & \textbf{0.9346} \\
    \midrule
    Abs. Improv. & 0.0770 & 0.1288 & 0.0449 & 0.3054 & 0.2157 & 0.1782 & 0.0403 & 0.0521 & 0.0494 & 0.1710 & 0.1389 & 0.1099\\
    Rel. Improv.(\%) & 8.71\% & 15.73\% & 26.87\% & 69.81\% & 31.85\% & 22.40\% & 4.38\% & 6.13\% & 36.35\% & 24.85\% & 18.04\% & 13.33\%\\
    \bottomrule
    \end{tabular}
    \end{adjustbox}
\end{table*}

\begin{table*}[t!]
    \centering
    \caption{Results of the ablation study. {G}: Introducing Graph Structure, {H}: Hybrid Prediction Loss, {R}: Graph Regularization,
     {S}: Weighed Cosine Similarity, {W}: Weighting Negative Feedback, {P}: Multi-Head Attentive Pooling.}
    \label{tab:Ablation}
    \begin{threeparttable}
    \begin{tabular}{cc@{\hspace{18pt}}c@{\hspace{18pt}}c@{\hspace{18pt}}c@{\hspace{18pt}}c@{\hspace{18pt}}c|c@{\hspace{18pt}}c@{\hspace{18pt}}c|c@{\hspace{18pt}}c@{\hspace{18pt}}c}
        \toprule
           \multirow{2}{*}{\textbf{Model}} & \multicolumn{6}{c|}{\textbf{Settings}}  & \multicolumn{3}{c|}{\textbf{Real-Time Detection}} &  \multicolumn{3}{c}{\textbf{Post-Hoc Detection}} \\
            \cmidrule{2-13}
           & \textbf{G} & \textbf{H} & \textbf{R} & \textbf{S} & \textbf{W} & \textbf{P} & \textbf{AUC}$\uparrow$ & \textbf{DR}$\uparrow$ & \textbf{FPR}$\downarrow$ & \textbf{AUC}$\uparrow$ & \textbf{DR}$\uparrow$ & \textbf{FPR}$\downarrow$ \\
        \midrule
        \textbf{\mname} & \Checkmark & \Checkmark & \Checkmark & \Checkmark & \Checkmark & \Checkmark & \textbf{0.9369} & \textbf{0.8869}  & \textbf{0.1411} & \textbf{0.9607} & \textbf{0.9478} & 0.1222\\
        \textbf{\mname -P} & \Checkmark & \Checkmark & \Checkmark & \Checkmark & \Checkmark & \XSolidBrush & 0.9253 & 0.8782  & 0.1535 & 0.9531 & 0.9059 & \textbf{0.1151} \\
        \textbf{\mname -P/W} & \Checkmark & \Checkmark & \Checkmark & \Checkmark & \XSolidBrush & \XSolidBrush & 0.8340 & 0.7809 & 0.2945 & 0.8898 & 0.8375 & 0.1721 \\
        \textbf{\mname -P/W/S} & \Checkmark & \Checkmark & \Checkmark & \XSolidBrush & \XSolidBrush & \XSolidBrush & 0.8325 & 0.8010 & 0.2979 & 0.8801 & 0.8298 & 0.1771 \\
        \textbf{\mname -P/W/S/R} & \Checkmark & \Checkmark & \XSolidBrush & \XSolidBrush & \XSolidBrush & \XSolidBrush & 0.8314 & 0.7983 & 0.3001 & 0.8653 & 0.8282 & 0.1939\\
        \textbf{\mname -P/W/S/R/H} & \Checkmark & \XSolidBrush & \XSolidBrush & \XSolidBrush & \XSolidBrush & \XSolidBrush & 0.8144 & 0.7750 & 0.2901 & 0.8631 & 0.8119 &  0.2159 \\
        \textbf{\mname -P/W/S/R/H/G} & \XSolidBrush & \XSolidBrush & \XSolidBrush & \XSolidBrush & \XSolidBrush & \XSolidBrush & 0.7469 & 0.7152 & 0.3767 & 0.8531 & 0.7891 & 0.2259 \\
        \bottomrule
    \end{tabular}
    \end{threeparttable}
\end{table*}

From \Cref{tab:Post-Hoc-Main} it can be seen that \mname achieves the best performance in terms of all six metrics on two datasets, with 96.07\% AUC, 94.78\% DR, 12.22\% FPR in CERT r4.2 and 96.05\% AUC, 90.24\% DR, 8.65\% FPR in CERT r5.2. The results demonstrate the effectiveness of \mname on post-hoc ITD, which can detect more anomalies with fewer false positives. Especially when the investigation budget is 15\% of the activities under test, the detection rate (i.e., DR@15\%) reaches 97.39\% and 93.46\% in CERT r4.2 and r5.2, respectively. Moreover, we can also observe that \mname surpasses all baselines by at least 4.49\% in FPR, which indicates a significant improvement in reducing the waste of the investigation budget.
Similar to the observations in \Cref{tab:Real-time-Main}, it can be seen that FMLP achieves suboptimal performance on the CERT r4.2 dataset, which might benefit from filtering out noise in the historical activity data.
Comparing \mname and the sequential models (i.e., DeepLog), it can be also seen that \mname significantly improves the performance in post-hoc ITD. 

Combining the results in \Cref{tab:Real-time-Main} and \Cref{tab:Post-Hoc-Main}, it can be concluded that \mname can be applied for both real-time ITD and post-hoc ITD, superior to all the state-of-the-art baselines with regards to all six metrics, especially in reducing false positives which might waste the manual investigation cost.
{Finally, we also record the inference time of \mname. The average inference time for real-time ITD and post-hoc ITD is 0.5023ms and 0.5675ms, respectively, which means \mname is a feasible solution for insider threat detection at the activity level.}

\subsection{Ablation Studies }

In this section, we conduct ablation studies to evaluate the effectiveness of each module in \mname.  
As the comparative experiments, we configure \mname by using LSTM as the sequence encoder and GCN as the graph neural network in \mname. The ablation study was conducted on the CERT r4.2 dataset. We implement \mname with the different settings as follows.

\vspace{0pt}
\begin{itemize}[leftmargin=11pt]
\item \textbf{\mname} represents the complete version of the model proposed in this paper. 
\item \textbf{\mname -P}. We further remove the multi-head attentive pooling operation mentioned in ~\Cref{Sec:Multi-Head Attentive Pooling} from the complete \mname, replacing it with the average pooling operation. 
\item \textbf{\mname -P/W}. On the basis of \mname -P, we replace the weight of negative feedback mentioned in \Cref{eq:hardweight} with 1, setting the same weights for normal and abnormal activities.
\item \textbf{\mname -P/W/S}. On the basis of \mname -P/W, we further remove the weighted cosine similarity metric function in \Cref{eq:weightedcosine}, and replace it with the cosine similarity.
\item \textbf{\mname -P/W/S/R}. On the basis of \mname -P/W/S, we further remove the graph regularization loss in \Cref{Sec:Graph Regularization}.
\item \textbf{\mname -P/W/S/R/H}. On the basis of \mname -P/W/S/R, we replace the hybrid prediction loss mentioned in \Cref{Sec:Hybrid Prediction Loss} with the standard cross entropy loss. 
\item \textbf{\mname -P/W/S/R/H/G}. On the basis of \mname -P/W/S/R/H, we remove the entire graph structure. At this point, the model is actually {a single sequence encoder} such as LSTM.
\end{itemize}
\vspace{0pt}

\Cref{tab:Ablation} shows the results of the ablation study.
From the experimental results, it can be observed that the introduction of each component leads to varying degrees of performance improvement in the model.
From the results of \mname -P/W/S/R/H/G, \mname -P/W/S/R, and \mname -P/W/S, it can be observed that the removal of these modules leads to a varying degree of performance decrease. This indicates that using our framework is crucial because it allows the model to break free from the limitations of sequence models and automatically discover relationships between activities located in different sequences, resulting in a significant improvement in model performance. 
From the results of \mname -P/W/S/R/H and \mname -P/W, it can be observed that the removal of these modules also leads to a significant decrease in performance. This indicates the use of hybrid prediction loss and weighting negative feedback is also very important. This is because 
they enable our framework to make full use of limited label information and enhance the discrimination ability for abnormal activities. Ultimately, in the real-time ITD task, our model improves the initial model's AUC from 0.7469 to 0.9369, increases DR from 0.7152 to 0.8869, and reduces FPR from 0.3767 to 0.1411. In the post-hoc ITD task, our model improves the initial model's AUC from 0.8531 to 0.9607, increases DR from 0.7891 to 0.9478, and reduces FPR from 0.225 to 0.122. These improvements are crucial for activity-level ITD.

\subsection{Parameter Analysis}
\label{Sec:Parameter Sensitivity}

\begin{figure}[t]
    \centering
    \subfloat[CERT r4.2]{
        \begin{minipage}[b]{0.45\linewidth}
            \includegraphics[width=\textwidth]{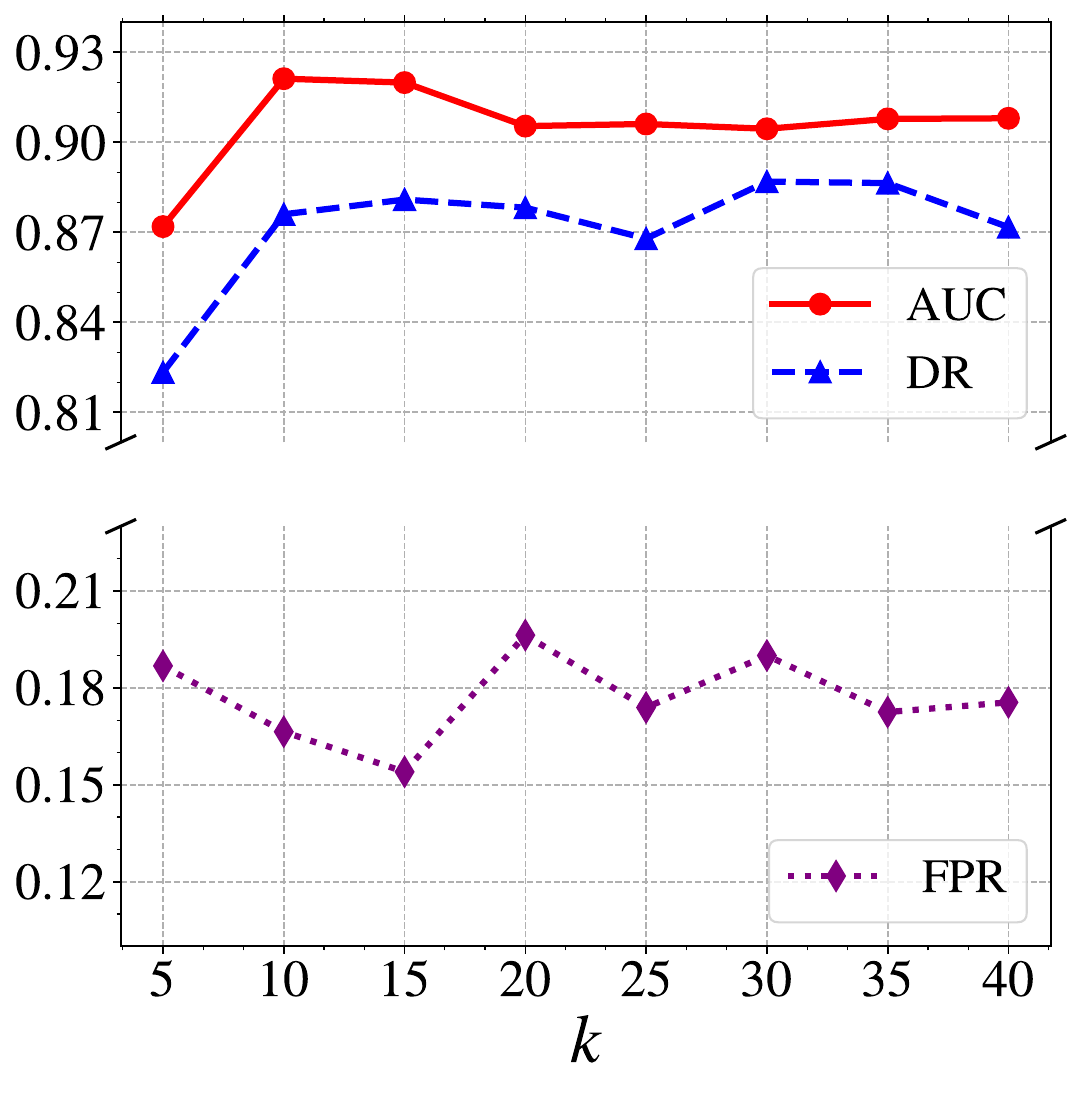} 
        \end{minipage}
    }
    \subfloat[CERT r5.2]{
        \begin{minipage}[b]{0.45\linewidth}
            \includegraphics[width=\textwidth]{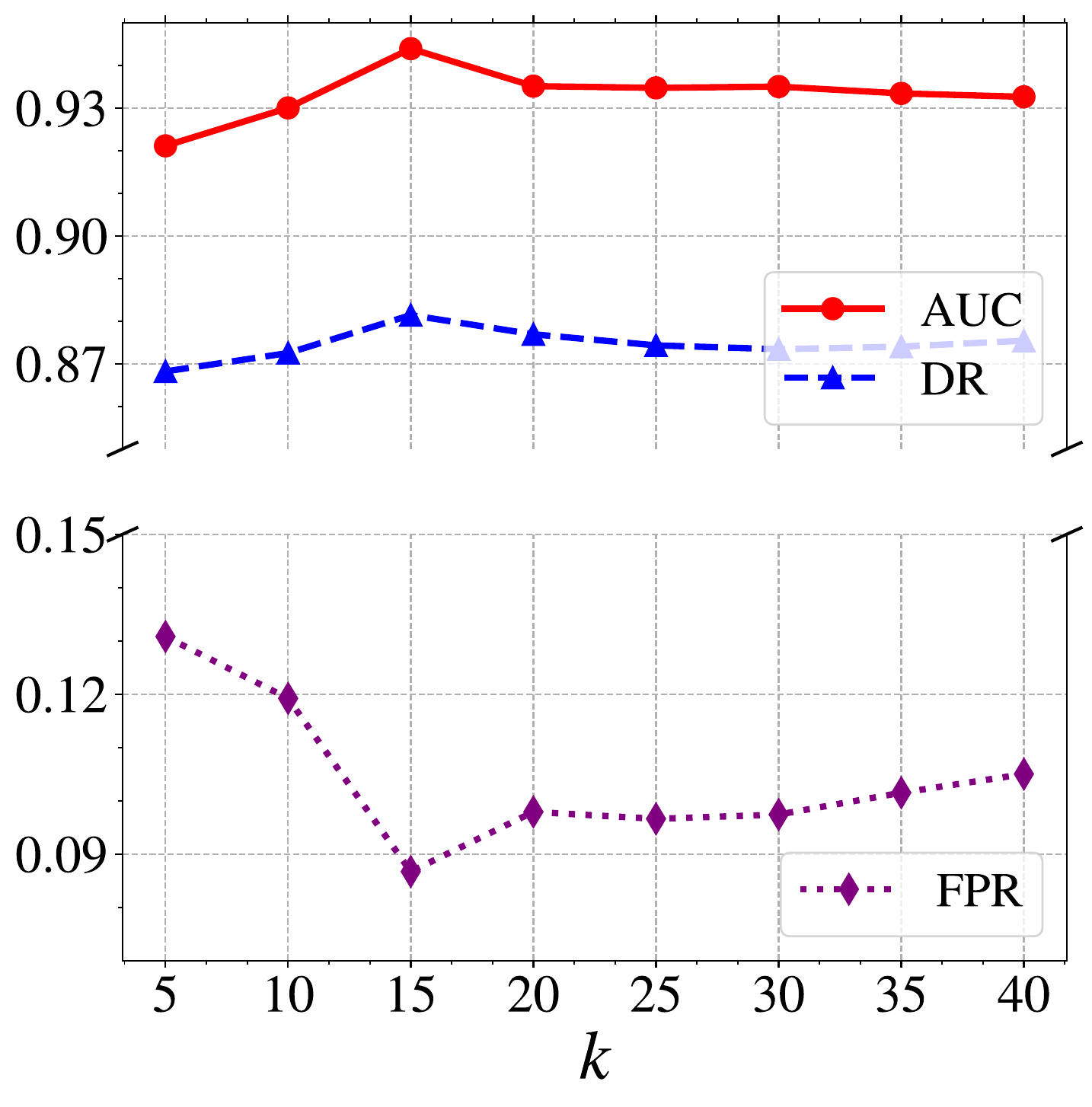} 
        \end{minipage}
    }
    
    \caption{Performance of \mname with different numbers of candidate neighbors $k$ obtained through retrieval}
    \label{fig:topk}
\end{figure}

\begin{figure}[t]
    \centering  
	\subfloat[CERT r4.2]{
		\begin{minipage}[b]{0.45\linewidth}
			\includegraphics[width=\textwidth]{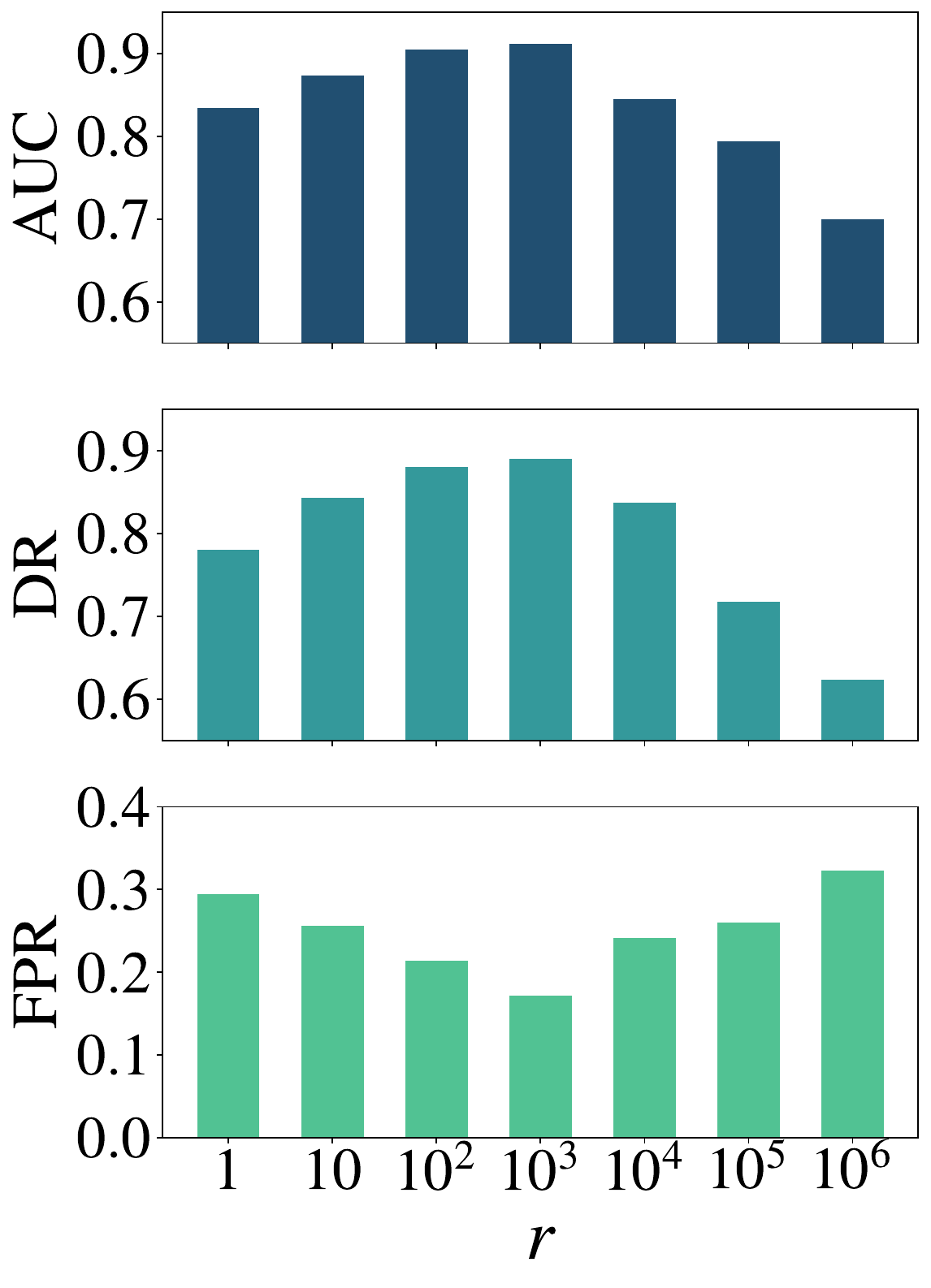} 
		\end{minipage}
		\label{fig:fig5a}
	}
    \subfloat[CERT r5.2]{
        \begin{minipage}[b]{0.45\linewidth}
        \includegraphics[width=\textwidth]{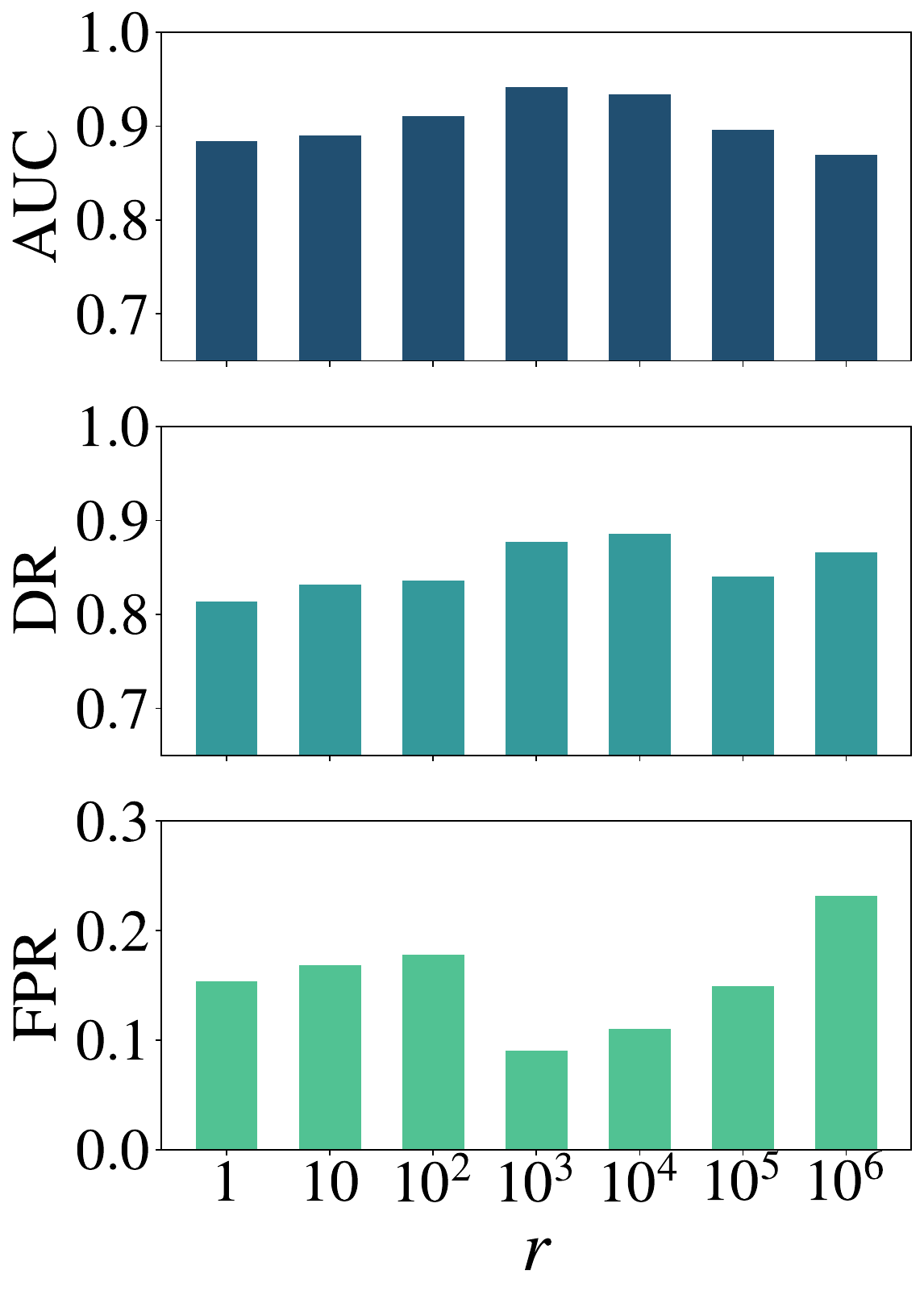}
        \end{minipage}
    \label{fig:fig5b}
    }
	\caption{Performance of \mname with different weights of negative feedback $r$ in the hybrid prediction loss}
	\label{fig:weight of hard negative sample}
\end{figure}

In this section, we analyze the influences of two key hyper-parameters of \mname, i.e., the number of candidate neighbors $k$ in \sihan{the \moduletwo module} and the weight of negative feedback $r$ in the hybrid prediction loss.
Specifically, we vary the number of candidate neighbors $k$ by 5, 10, 15, 20, 25, 30, and 40. \Cref{fig:topk} illustrates the changes of AUC, DR, and FPR as $k$ increases. 

On both datasets, it can be observed that as $k$ increases, AUC first increases rapidly and reaches the peak when $k=10$, and then decreases slowly. Similarly, when we increase $k$, DR first increases and then fluctuates slightly. In contrast to AUC and DR, as $k$ increases, FPR first drops rapidly, and then it increases with fluctuations. 

In summary, in terms of AUC, \mname achieves the best performance when $k$ is set to 10. In terms of FPR, \mname achieves the best performance when $k$ is 15 and the FPR is the lowest. 
Moreover, it can also be observed that in many cases, when we increase $k$, both DR and FPR increase at the same time. A possible reason is that by considering more neighbors, the aggregation activities are more diverse, which makes the model detect more abnormal activities, but also introduces more noise. Overall, the changes in performance are not significant, which might be due to the noise-resistant capability provided by the design of \mname such as the graph regularization constraints and weighted cosine metric.

For the weight of negative feedback $r$ in the hybrid prediction loss, we vary it by $1$, $10$, $10^2$, $10^3$, $10^4$, $10^5$, and $10^6$. \Cref{fig:weight of hard negative sample} shows the results of AUC, DR, and FPR with different weights $r$. Overall, it can be seen that the influence of the weight $r$ exhibits similar trends in the two datasets.
As the negative sample weight $r$ increases, AUC and DR first increase and reach the peak. After the peak, AUC and DR decrease when $r$ increases. In contrast, when we increase $r$, FPR decreases at first, and achieves the lowest value when AUC and DR reach the peak. After that, FPR increases slowly. Specifically, on the CERT r4.2 dataset, \mname achieves the best performance when $r$ is set to $10^3$, which is very close to the imbalance ratio of CERT r4.2 (i.e., 1,048).
A possible explanation is that the model with $r$ set by the imbalance ratio can pay appropriate attention to the negative feedback, which improves the detection ability for abnormal activities, and prevents the model from yielding to the majority of samples, which are normal activities.
On the CERT r5.2 dataset, we obtain a similar observation. Specifically, when $r$ is $10^3$, AUC is the highest and FPR is the lowest among all values. When $r$ is $10^4$, DR achieves the highest value. The observation is consistent with the imbalance ratio of CERT r5.2 (2,645), which is between $10^3$ and $10^4$.

\subsection{Compatibility Analysis}
\label{Sec:Compatibility Analysis}

\begin{table}[t]
    \centering
    \caption{
Performance of different combinations of models}
    \label{tab:Architecture}
    \begin{tabular}{c|cc|ccc}
        \toprule
        & \textbf{Dataset} & \textbf{Architecture} & \textbf{AUC} $\uparrow$ & \textbf{DR} $\uparrow$ & \textbf{FPR} $\downarrow$ \\
        
        \midrule
        \multirow{12}{*}{\begin{sideways}\textbf{Real-Time}\end{sideways}}& \multirow{6}{*}{r4.2} & GRU+GCN & 0.9313 & 0.8983 & 0.1579 \\
        & & GRU+GAT & 0.9298 & 0.8826 & 0.1499 \\
        & & LSTM+GCN & \textbf{0.9369} & 0.8875  & \textbf{0.1411} \\
        & & LSTM+GAT & 0.9309 & \textbf{0.9141} & 0.1447\\
        & & Transformer+GCN & 0.9086 & 0.8217 & 0.1765 \\
        & & Transformer+GAT & 0.9060 & 0.8385 & 0.1681 \\
        \cmidrule(r){2-6}
        & \multirow{6}{*}{r5.2} & GRU+GCN & 0.9420 & 0.8599 & 0.0923\\
        & & GRU+GAT & 0.9434 & 0.8794 & 0.1006\\
        & & LSTM+GCN & \textbf{0.9439} & 0.8814 & 0.0867 \\
        & & LSTM+GAT & 0.9426 & \textbf{0.8829} & 0.1000 \\
        & & Transformer+GCN & 0.9329 & 0.8700 & 0.1098 \\
        & & Transformer+GAT & 0.9349 & 0.8513 & 0.0831 \\
        \midrule
        \multirow{12}{*}{\begin{sideways}\textbf{Post-Hoc}\end{sideways}}&  \multirow{6}{*}{r4.2} & GRU+GCN & \textbf{0.9610} & 0.9429 & 0.1262 \\
        & & GRU+GAT &  0.9590 & 0.9423 & 0.1206 \\
        & & LSTM+GCN & 0.9607 & \textbf{0.9478} & 0.1222 \\
        & & LSTM+GAT & 0.9578 & 0.9347 & \textbf{0.1199} \\
        & & Transformer+GCN & 0.9306 & 0.8934 & 0.1407\\
        & & Transformer+GAT & 0.9391 & 0.9092 & 0.1405 \\
        \cmidrule(r){2-6}
        & \multirow{6}{*}{r5.2} & GRU+GCN &0.9602 & \textbf{0.9035} &  0.0816 \\
        & & GRU+GAT & 0.9575 & 0.8952 & 0.0779\\
        & & LSTM+GCN & 0.9605 & 0.9024 & 0.0865  \\
        & & LSTM+GAT & \textbf{0.9607} & 0.8984 & \textbf{0.0767} \\
        & & Transformer+GCN & 0.9410 & 0.8791 & 0.1013 \\
        & & Transformer+GAT &0.9446 & 0.8794 & 0.0966  \\
        \bottomrule
    \end{tabular}
\end{table}

\begin{figure}[t]
    \centering  
    \subfloat[\mname \ on Real-Time ITD]{
        \includegraphics[width=0.49\linewidth]{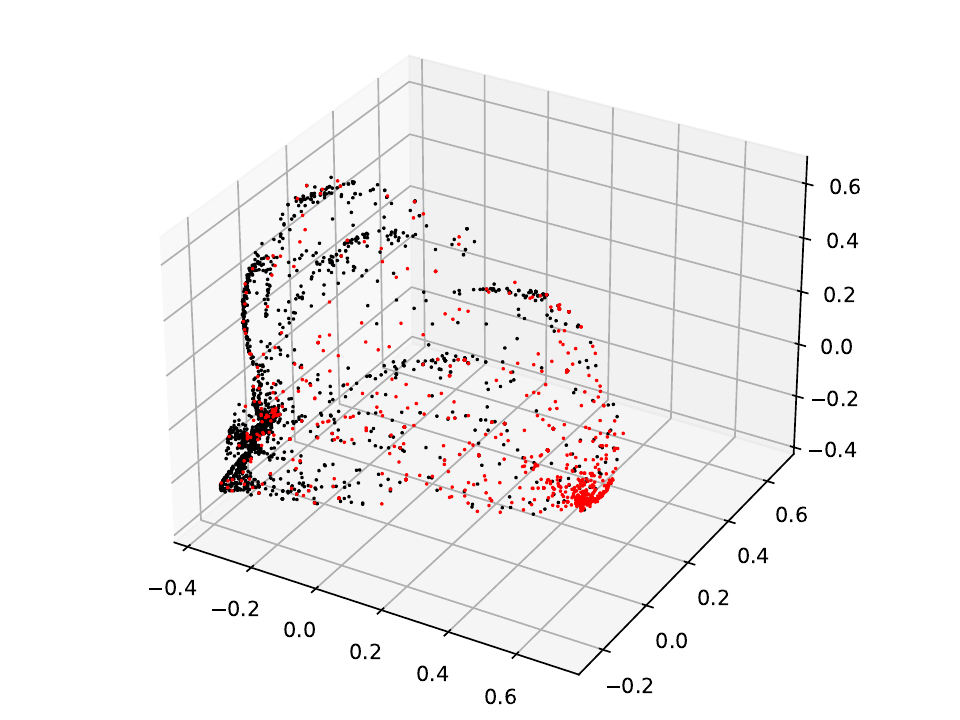} 
        \label{fig:OurRTVis}
    }
    \subfloat[DeepLog on Real-Time ITD]{
        \includegraphics[width=0.49\linewidth]{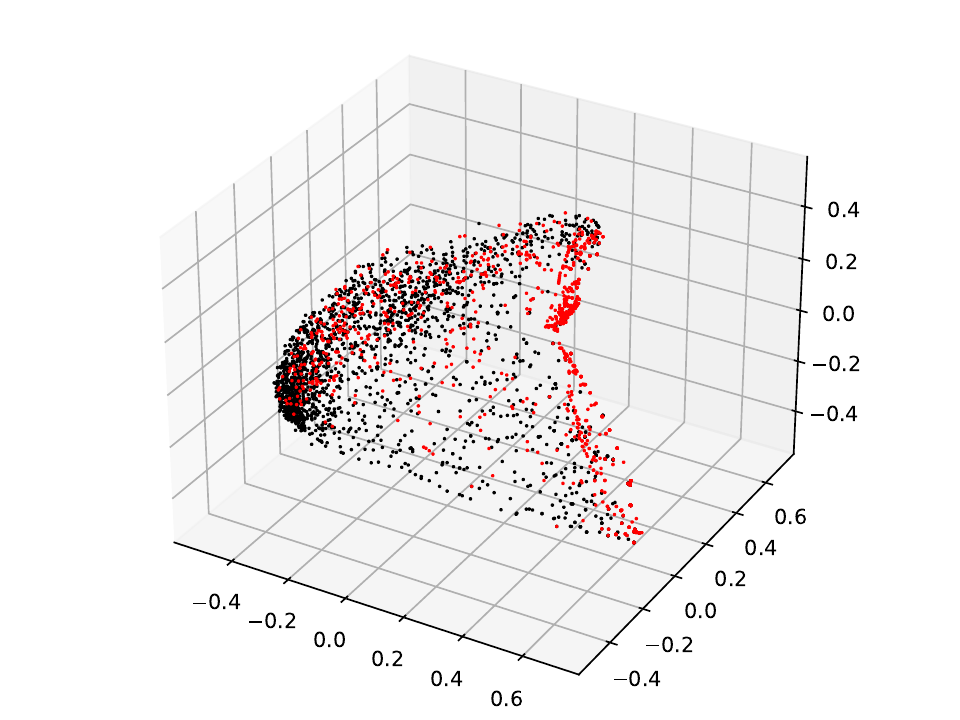} 
        \label{fig:DeepLogRTVis}
    } 
    \\
    \subfloat[\mname \ on Post-Hoc ITD]{
        \includegraphics[width=0.49\linewidth]{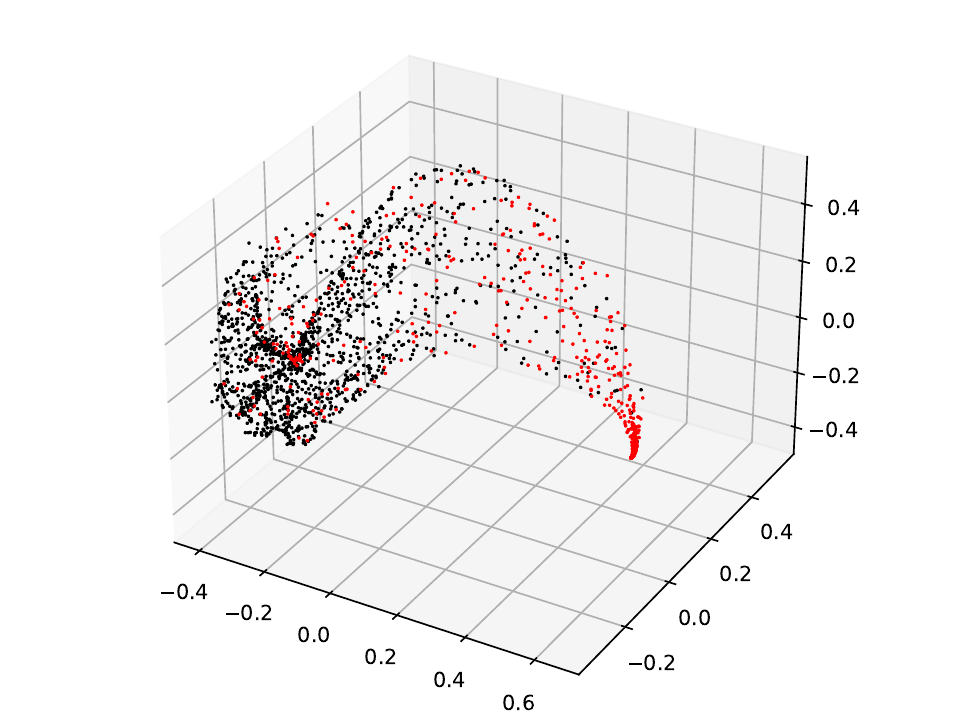}
        \label{fig:OurPHVis}
    }
    \subfloat[DeepLog on Post-Hoc ITD]{
        \includegraphics[width=0.49\linewidth]{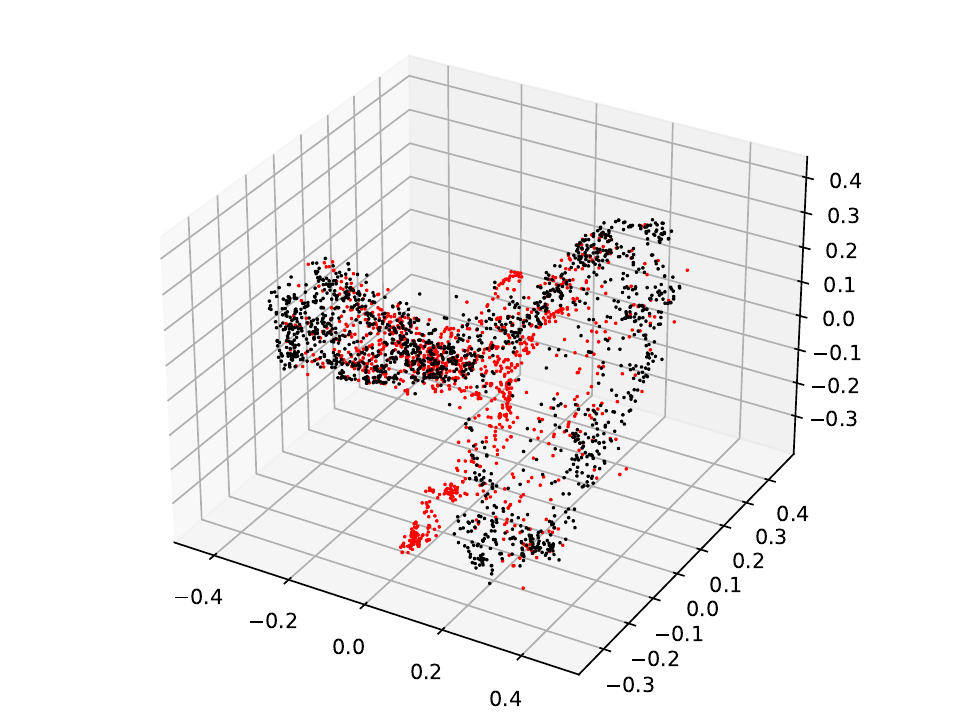}
        \label{fig:DeepLogPHVis}
    }
    \caption{Visualization of activity vectors. The black and red dots are the vectors of normal and abnormal activities, respectively.}
    \label{fig:Vis}
\end{figure}

Since the compatability of the proposed framework is significant for its practical use, we realize different implementations of \mname to verify the proposed framework for both real-time ITD and post-hoc ITD. Specifically, in the \moduleone module, we implement \mname with three representative sequence encoders, i.e., LSTM, GRU, and Transformer. For post-hoc detection, we configure LSTM and GRU as bidirectional to utilize context before and after a timestamp simultaneously. In the \modulethree module, we implement \mname with two representative graph neural networks, i.e., GCN and GAT. To conduct a fair comparison, the rest of the framework is the same for all models, and we test the implementations on two datasets for both real-time ITD and post-hoc ITD.

\Cref{tab:Architecture} presents the experimental results of different implementations of \mname.
First, it can be observed that all combinations perform well in both real-time ITD and post-hoc ITD on two datasets, surpassing the performance of all the baselines as shown in \Cref{tab:Real-time-Main} and \Cref{tab:Post-Hoc-Main}. The results demonstrate the effectiveness and compatibility of the proposed framework, reflecting that \mname can incorporate the temporal dependencies and learn the relationships between user activities across different sequences simultaneously.

Among all the implementations of the proposed framework, it can be observed from \Cref{tab:Architecture} that the use of LSTM and GCN leads to slightly better performance than other models.  
Regarding the GNN module, using GAT in \mname \ often leads to a slight decrease in AUC performance. However, in many cases, using GAT exhibits superior DR. The potential reason for this is that GAT may attempt to focus more on specific neighbors, breaking the limitations of graph structure learning.
Nonetheless, breaking these limitations is not always beneficial, as it may introduce some noise. However, at times, it can also yield unexpected effects.
Another interesting observation is that, while ~\Cref{tab:Real-time-Main} and ~\Cref{tab:Post-Hoc-Main} show that the more complex and global sequence encoder, Transformer, generally outperforms the other three recurrent sequence encoders (RNN, GRU, and LSTM) when used independently, especially in terms of FPR. 
However, when used as a sequence encoder as part of our \mname framework, although its performance is also improved compared to using Transformer independently, and surpasses all other state-of-the-art methods, it is not as good as setting the sequence encoder of \mname \  directly to LSTM. A possible reason is that the representations generated by Transformer exhibit significant anisotropy in the vector space ~\cite{DBLP:conf/emnlp/Ethayarajh19, DBLP:conf/iclr/GaoHTQWL19}, occupying a narrow cone-like structure. As a result, each output representation tends to be similar, which hinders the learning of graph structure.

\subsection{Visualization}
{To further illustrate the difference between \mname and DeepLog, which are both implemented based on LSTM, we exploit Principal Component Analysis (PCA) to project the user activity vectors learned by two methods into a three-dimensional space. \Cref{fig:Vis} shows the visualization of the activity vectors for both real-time ITD and post-hoc ITD,
where the black and red points represent the vectors of normal activities and abnormal activities, respectively. It can be observed that in the latent space, \mname achieves better distinguishment between normal and abnormal activities than DeepLog. The visualization demonstrates the effectiveness of learning the relationships between user activities across different sequences by graph neural networks.}

\section{Related Work} \label{sec:rw}

In this section, we summarize the related work in the field of insider threat detection. Existing works can be grouped into three categories, i.e., the feature engineering-based methods, the sequence-based methods, and the graph-based methods. 

\subsection{Feature Engineering-based methods}
{The first group of studies relies on feature engineering to detect insider threats~\cite{legg2015caught,le2020analyzing,le2021anomaly,liu2018anomaly,yuan2021time}. Specifically, they extract the features for a user or a time period, such as the number of websites accessed on a shared PC, the number of sent emails, and the average size of email attachments.}
Based on the extracted features, a {machine learning-based anomaly detection model} is trained, such as logistic regression, random forest, neural network, XGBoost, autoencoder, and isolation forest.
{These studies detect insider threats at the user, weekly, daily, or session levels. Unlike these studies, in this paper, we conduct the first study on activity-level real-time ITD.}

\subsection{Sequence-based Methods.} 
Since feature engineering-based methods require extensive domain expert knowledge to select appropriate features for feature extraction, much effort has been made to automatically learn the representations of user behaviors. 
In recent years, deep learning techniques have gained much attention, and many works have introduced deep learning techniques for ITD.
A natural practice is to aggregate user activities into an activity sequence and use sequence models in the natural language processing (NLP) field for anomaly detection ~\cite{vinay2022contrastive,tuor2017deep,yuan2020few,yuan2019insider,lu2019insider,DBLP:conf/iscc/HuangZLLWY21}.
With sequence models, these studies can incorporate the temporal dependencies between user activities.
Specifically, Yuan et al.~\cite{yuan2019insider} proposed a model that combines temporal point processes and recurrent neural networks for sequence-level ITD.
After that, their follow-up work treated user behaviors as a sequence of activities and used few-shot learning to detect sequence-level insider threats~\cite{yuan2020few}.
Huang et al.~\cite{DBLP:conf/iscc/HuangZLLWY21} pre-trained a language model BERT~\cite{devlin2018bert} on the historical activity data and used a bidirectional LSTM for sequence-level detection.
{Tuor et al.~\cite{tuor2017deep} first extracted features for each user daily, and then fed the historical feature vectors into an LSTM to predict the feature vector of the next day for daily-level ITD.}

\subsection{Graph-based methods.} 
Recently, to incorporate more relationships between users and activities, graph neural networks have been widely used in the field of ITD \cite{DBLP:journals/tifs/LiLJLYGY23,zheng2022insider,jiang2019anomaly,hong2022graph,DBLP:conf/ccs/LiuWZJXM19}.
Specifically, Jiang et al.~\cite{jiang2019anomaly} considered that user relationships can provide powerful information for detecting abnormal users. 
They modeled the relationships between users within an organization as a graph using email communication and user-based features, applying graph convolutional networks to detect insiders.
Li et al.~\cite{DBLP:journals/tifs/LiLJLYGY23} converted user features and the user interaction structure into a heterogeneous graph and then used a dual-domain graph convolutional network to detect anomalous users.
{Liu et al.~\cite{DBLP:conf/ccs/LiuWZJXM19} represented user activities with nodes and designed several heuristic rules for graph construction. 
Finally, they constructed a heterogeneous graph, applied graph embedding algorithms on the graph, and utilized a clustering algorithm to detect anomalous activities.
This is the only study that has focused on activity-level detection. However, it only considered post-hoc ITD and relied on heuristic rules designed by experts to construct graphs.}
In this paper, we employ graph structure learning to learn the user activity graph adaptively, avoiding the bias introduced by manual graph construction.

\section{Conclusion} \label{sec:con}
In this paper, we take the first step towards real-time ITD at the activity level. We present a fine-grained and efficient framework \mname, which can be applied for real-time ITD and post-hoc ITD with slight modifications. 
It leverages graph structure learning to autonomously learn the user activity graph without manual intervention, incorporating both the temporal dependencies and the relationships between user activities across different activity sequences.
Furthermore, to mitigate the data imbalance issue in ITD, we also propose a novel hybrid prediction loss that integrates self-supervision signals from normal activities and supervision signals from abnormal activities into a unified loss. 
Extensive experiments demonstrate the effectiveness of \mname, superior to 9 state-of-the-art methods for real-time ITD and 8 competitive baselines for post-hoc ITD. We also conduct the ablation study and parameter analysis to measure the effectiveness of each component and hyper-parameters.
One future plan is to integrate more effective sequence encoders and GNNs to improve performance and time efficiency.  Moreover, since labeling abnormal samples requires much effort, we plan to design an interactive framework for anomaly detection.

\normalem
\bibliographystyle{IEEEtran}
\bibliography{ref}

\end{document}